\documentclass[referee]{raa}
\usepackage{graphicx,times}
\usepackage{natbib}
\usepackage{amssymb,amsmath}
\usepackage{float}
\usepackage{rotating}
\usepackage{tabularx} 
\usepackage{makecell} 
\usepackage{array}
\usepackage{multirow}
\usepackage{setspace}
\usepackage{longtable}
\usepackage{footnote}
\usepackage{tablefootnote}
\usepackage{ragged2e}
\usepackage[bottom]{footmisc}
\bibpunct{(}{)}{;}{a}{}{,}

\usepackage[pagebackref=true]{hyperref}

\begin{document}

   \title{A catalogue and statistical analysis for magnetic stars}

 \volnopage{ {\bf 20XX} Vol.\ {\bf X} No. {\bf XX}, 000--000}
   \setcounter{page}{1}

   \author{Abdurepqet Rustem
   \inst{1,2}, Guo-Liang L{\scriptsize$\ddot{\rm U}$}\inst{2,1}, Jin-Zhong Liu\inst{2}, Chun-Hua Zhu
      \inst{1}, Yu-Zhang\inst{2}, Dong-Xiang Shen\inst{1}, Yu-Hao Zhang\inst{1}, Xiao-Long He\inst{2}
   }

   \institute{ School of Physical Science and Technology, Xinjiang University, Urumqi 830046, China;\\
        \and
             Xinjiang Astronomical Observatory, Chinese Academy of Sciences, Urumqi 830011, China;{\it guolianglv@xao.ac.cn};{\it liujinzh@xao.ac.cn}\\
\vs \no
{\small Received 20XX Month Day; accepted 20XX Month Day}
}

\abstract{
Magnetic fields are significant in the structure and evolution of stars.
We present a comprehensive catalogue of 1784 known magnetic stars, detailing their identifications, HD numbers, precise locations, spectral types, and averaged quadratic effective magnetic fields among other important information.
The group comprises 177 O-type stars, 551 B-type stars, 520 A-type stars, 91 F-type stars, 53 G-type stars, 61 K-type stars, 31 M-type stars, and an additional 300 stars whose spectral classification remains indeterminate.
Our analysis examines the statistical properties of these magnetic stars.
The relative integrated distribution function and number distribution function for all magnetic stars of the same spectral type can be effectively approximated using an exponential function of the averaged quadratic effective magnetic field.
The analysis further reveals that A and B-type stars possess the strongest mean magnetic fields, indicating an easier detection of their magnetic fields.	
}

   \authorrunning{Abdurepqet Rustem et al. }            
   \titlerunning{A catalogue and statistical analysis for magnetic stars}  
   \maketitle

%
\section{Introduction}           
\label{Sect1}

It is a widely acknowledged fact that the structure and evolution of stars are primarily determined by their mass and metallicity \citep{1990sse..book.....K}.
Gradually, the significance of magnetic field effects in stars is being recognized \citep{2009ARA&A..47..333D}.
The study of magnetic fields in stars started with Hale's detection of magnetic fields in sunspots back in 1908 \citep{1908ApJ....28..315H}.
\cite{1958ApJS....3..141B} compiled the first-ever catalogue of magnetic fields by gathering an enormous number of observations made over 11 years using the Zeeman Effect analyzer mounted on 100-inch and 200-inch telescopes.
\cite{2003A&A...407..631B,2009MNRAS.394.1338B} amassed a substantial volume of magnetic field observations of main-sequence stars and giants based
\newpage
\justify
on CP stars. After more than a century of hard work, it has been found that low-mass stars (solar-like or cool stars) typically exhibit some form of magnetic activity, such as dark spots, prominences, flares, mainly occurring on their outer layer, and that the magnetic energy is a result of either convective or rotational energy \citep{2009ARA&A..47..333D,2010LRSP....7....3C}.

Compared to low-mass stars, approximately 7\% of massive stars (O or B type) have been observed to possess magnetic fields \citep{1992A&ARv...4...35L}, and their magnetic fields tend to be stronger and have simpler structures \citep[e. g.,][]{2009ARA&A..47..333D}.
Since 2014, several large surveys such as the B fields in OB Stars \citep{2014Msngr.157...27M}, the Magnetism in Massive Stars project \citep{2016MNRAS.456....2W}, and the Large Impact of Magnetic Fields on the Evolution of Hot stars project \citep{2018MNRAS.475.1521M}, aimed at measuring the magnetic fields of massive stars, have been conducted.

Indeed, these surveys have been tremendously valuable in uncovering the properties of magnetic massive stars and identifying hundreds of new ones \citep{2017MNRAS.465.2432G,2018pas8.conf..146S,2019MNRAS.490.4154S,2019MNRAS.482.3950S,2023RAA....23a5002S,2023Galax..11...40K}.
However, to date, there is still no complete catalogue for these magnetic massive stars that supersedes the one created by \cite{2009MNRAS.394.1338B}.

In the present paper, we have meticulously assembled a catalogue that showcases all of the publicly available magnetic stars.
This catalogue provides comprehensive information regarding the mean longitudinal (or effective) magnetic fields associated with these celestial bodies.
In addition, extensive observations and recorded sources are also included to support the measurement of the aforementioned magnetic fields, affording readers an opportunity to thoroughly scrutinize and analyze our findings.
The present paper is divided into three sections.
Section \ref{Sect2} entails detailed discourse relating to the compilation process for the magnetic stars catalogue.
Section \ref{Sect3} offers a statistical analysis of the magnetic stars described in our catalogue. 
Finally, Section \ref{Sect4} contains a discussion and concluding remarks.


\section{Catalogue of Magnetic Stars}
\label{Sect2}
Drawing from an extensive collection of past stellar magnetic field observation records and utilizing the Simbad database, we have identified 1,784 magnetic stars.
This section will comprehensively depict the data, selection criteria, and methodological approach employed to create our exhaustive catalogue of magnetic stars.

\subsection{Data}
\label{Sect2.1}
In our research, we have gathered comprehensive data on the mean longitudinal magnetic fields of magnetic stars, dating back to the initial discovery of sunspot magnetic fields up until the present day.
This dataset encompasses a total of 1,784 stars and includes 15,581 individual measurements of the average longitudinal magnetic field, all of which were derived from 223 pertinent published papers.
Each measurement represents the magnitude of the star’s magnetic field measured at a given time.
Of course, stellar magnetic fields vary dipole-wise.
Therefore, these mean longitudinal magnetic fields do not accurately represent the combined or global magnetic field size of the star.

\begin{figure}
	\centering
	\includegraphics[width=12.0cm, angle=0, scale=0.9]{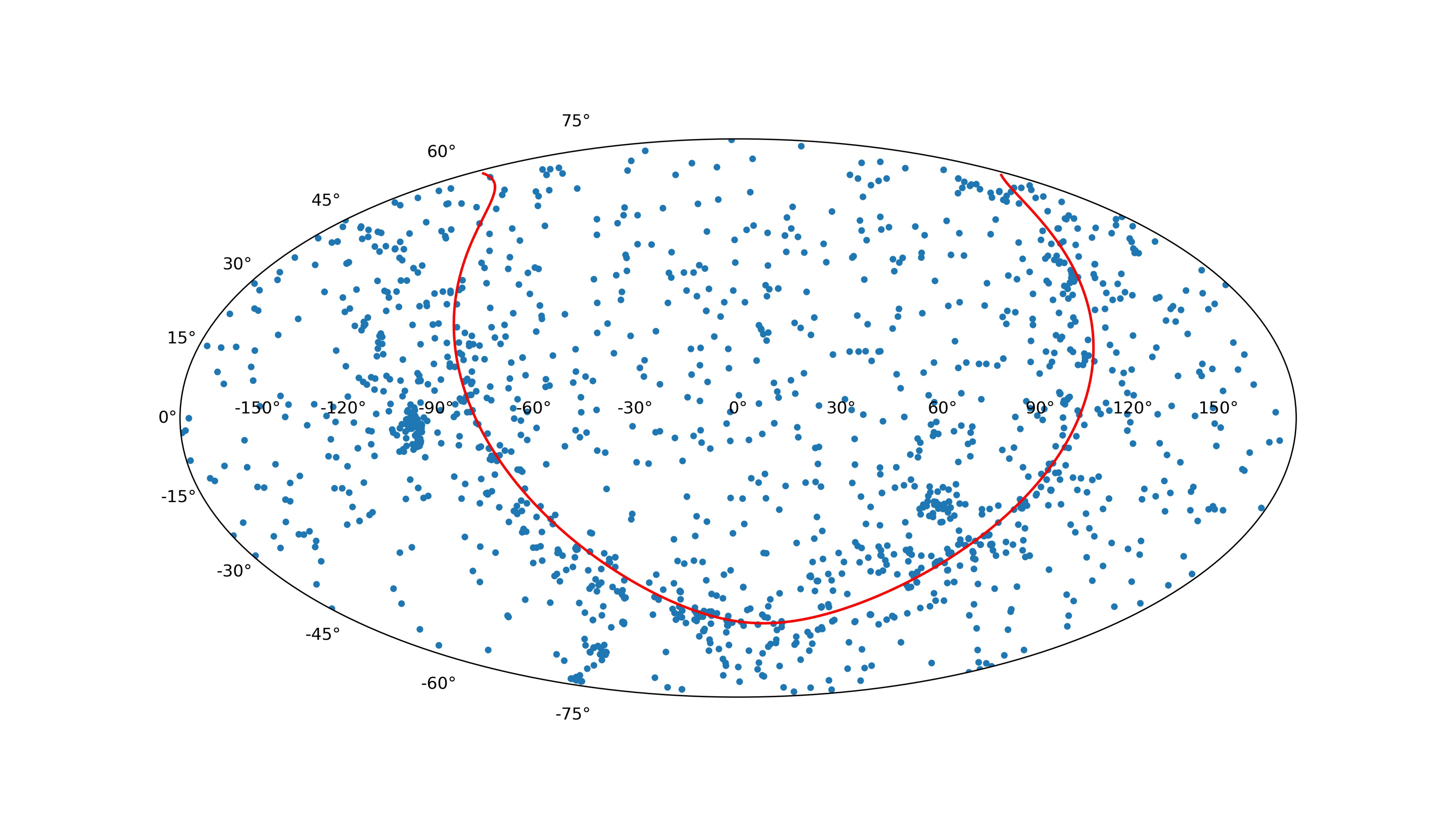}
	\caption{The positions of 1784 magnetic stars on the celestial sphere. The red solid line denotes the position of the Galactic plane. The right ascension and declination coordinates for the stars come from the Simbad database.}
	\label{Fig1}
\end{figure}

\newpage
\justify
Figure \ref{Fig1}, which is based on the right ascension and declination coordinates of the 1,784 magnetic stars, illustrates their positions on the celestial sphere.
These positions appear to follow a pattern characterized by an outward dispersion from the galactic plane.
However, it is essential to keep in mind that this distribution may be affected by observational bias introduced by the selection criteria used during the underlying Magnetic Stars Survey project.
While our research indicates an association between the spatial distribution of magnetic stars and their positioning away from the galactic plane, it does not necessarily suggest a specific clustering of magnetic stars within that region of the galaxy.
Hence, further investigations are required to better understand the nature and interaction of stellar magnetic fields with their surroundings.

\begin{figure}[t]
	\centering
	\includegraphics[width=12.0cm, angle=0, scale=0.8]{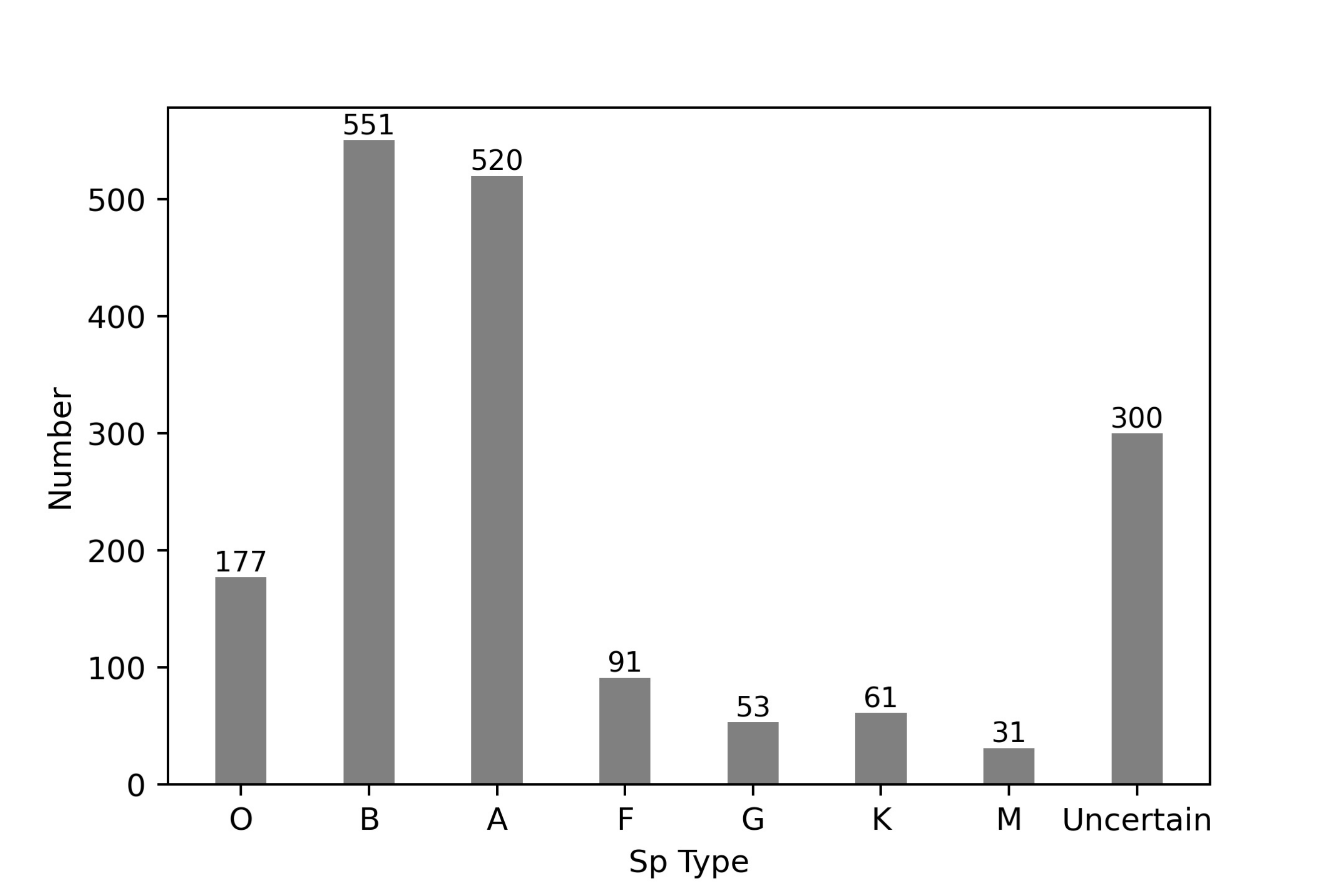}
	\caption{A bar chart depicting the number distribution of magnetic star spectral types.}
	\label{Fig2}
\end{figure}

Thanks to the utilization of advanced observational equipment and techniques, the number of magnetic stars observed in this paper has significantly increased by approximately 1.5 times compared to the observations made by \cite{2009MNRAS.394.1338B}.
We used the Simbad database and other resources with similar functions to determine the spectral classification of the magnetic stars in our sample.
Based on spectral analysis, the 1784 magnetic stars consist of 177 O-type, 551 B-type, 520 A-type, 91 F-type, 53 G-type, 61 K-type, 31 M-type stars, and 300 unknown type stars ; this information is presented in Figure \ref{Fig2}.
The surveys conducted for measuring magnetic fields of massive stars have led to a rapid increase in the number of magnetic massive stars, with early-type (O, B) stars accounting for approximately 50\% of all known spectral-type magnetic stars.
\cite{2009ARA&A..47..333D}, \cite{2012ASPC..464..405W,2016MNRAS.456....2W} explain that these early-type stars possess large-scale, simple, and stable magnetic structures, which make them unique and ideal for studying the principles of stellar magnetism.
Hence, these magnetic massive OB stars provide a distinctive environment to explore the principles of stellar magnetism.

\subsection{Selection criterions}
\label{Sect2.2}
Generally, the precise measurement of stellar magnetic fields is quite challenging.
Different equipment or
\newpage
\justify
techniques may result in varying magnetic fields estimates for a single star.
In our study, we have determined 15,581 magnetic strength measurements for 1,784 magnetic stars.
To ensure data accuracy and completeness while accounting for observational errors and time constraints, the selected stellar magnetic field measurements were based on three criteria:

The first selection criterion we used involved eliminating magnetic field data that was incomplete or outdated.
In particular, \cite{1983A&AS...53..119D} compiled a comprehensive catalogue of magnetic stars based on observations made between 1958 and 1983.
These observations were drawn from various literature sources published during that timeframe, which included many incomplete data points.
To differentiate between recent and older observations of stellar magnetic fields, we used this catalogue as a reference point.
If we collected new magnetic field data for stars already included in this catalogue, we replaced the earlier data with the most recent data.
Moreover, we removed any observations of stellar magnetic field strength exceeding 50,000 $G$ from the catalogue.
However, if an observation of magnetic field strength in the catalogue was unique, we retained that data.

The second criterion for selection entailed the evaluation of measurement errors present in the observed magnetic field values.
In the course of our data collection process, we discovered that some magnetic field measurement errors were not documented in relevant literature sources and catalogues.
Alternatively, some catalogues may have become too outdated to provide a comprehensive record of magnetic field observation errors.
Consequently, we identified 25 stars in our magnetic star catalogue with no documented magnetic field errors such as HD 166 and HD 13480, which only have recorded magnetic field magnitudes without accompanying errors.
With regard to whether or not to include such unerrored stellar measurements in the entire catalogue, only unique observations are retained while non-unique ones are omitted.

The third selection criterion pertains to the limitations of previous observations on stellar magnetic field intensities.
Some literature sources are too outdated to obtain records of observations or catalogues contai-
\newpage
\justify
ning magnetic field data points, necessitating their exclusion from our data set.
Fortunately, these limitations were relatively infrequent in occurrence.

\subsection{Methods}
\label{Sect2.3}
As far back as 2003, \cite{2003A&A...407..631B} introduced a catalogue and methodology for obtaining the averaged quadratic effective magnetic field $\left \langle B_{\rm e}\right \rangle$ of 596 main sequence and giant stars.
It is widely recognized that longitudinal magnetic field values display periodic variation based on rotational phase in most stars, resulting in polarity changes from positive to negative or even zero.
This issue can be effectively resolved through the utilization of the aforementioned method introduced by \cite{2003A&A...407..631B}, which computes the averaged quadratic effective magnetic field.
Building upon the work of \cite{2003A&A...407..631B}, we have also computed $\left \langle B_{\rm e}\right \rangle$ for 1784 magnetic stars in our present study using their equation:

\begin{equation} \label{eq1}
	\left \langle B_{\rm e}\right \rangle=\left(\frac{1}{n}\sum_{i=1}^nB_{{\rm e}i}^2\right)^{1/2},
\end{equation}
\begin{equation} \label{eq2}
	\left \langle\sigma_{\rm e}\right \rangle=\left(\frac{1}{n}\sum_{i=1}^n\sigma_{{\rm e}i}^2\right)^{1/2},
\end{equation}
where $B_{{\rm e}i}$, $i$ indicates the $i$th measurement of the mean longitudinal magnetic fields, with $n$ being the total number of observations for a given star.
The variable $\sigma_{{\rm e}i}$ represents the standard error of $B_{{\rm e}i}$, and $\left \langle\sigma_{\rm e}\right \rangle$ denotes the root-mean-square (rms) standard error of $\left \langle B_{\rm e}\right \rangle$.
Given that data collection occurs over an extended period using various equipment and observers, variations in the mean longitudinal magnetic fields may occur, with some stars exhibiting significant fluctuations.
To obtain a more precise representation of a stellar magnetic field, we rely on the averaged quadratic effective magnetic field.
The values derived from our rms calculation provide a more consistent depiction of the magnetic field's actual magnitude as it captures the star's global magnetic field rather than merely limited observations at a particular time scale.

Concurrently, \cite{2003A&A...407..631B} proposed the utilization of statistical deviations $\chi^2/n$ to accurately assess the dependability of a series of $B_{{\rm e}i}$ measurements.
Specifically, $\chi^2/n$ can be expressed as

\begin{equation} \label{eq3}
	\chi^2/n=\frac{1}{n}\sum_{i=1}^n\frac{B_{{\rm e}i}^2}{\sigma_{{\rm e}i}^2}
\end{equation}
where $n$ is the number of magnetic field measurements of the star.
The chi-square test is a common approach employed to evaluate hypotheses for counting data.
Its value reflects the difference between theoretical predictions and actual observations.
The value obtained from the test directly illustrates the discrepancy between the anticipated theoretical results and the factual observations.
In particular, a higher value of the test statistic signifies a more significant deviation from the expected model.

\subsection{Description of the catalogue}
\label{Sect2.4}
We have developed a comprehensive catalog of observed 1784 magnetic stars, which includes detailed information such as their names, HD numbers, positions, spectral types, averaged quadratic effective magnetic
\newpage
\justify
fields and corresponding calculation errors, chi-square values, calculation methods, observational counts for each magnetic star, and references.
As an example, Table \ref{Table1} shows the tiny part of the whole catalogue which is given in the appendix.
The catalogue records binary, double and multiple stars, with their respective HD numbers specified according to data sources, such as HD162305North, HD162305South, HD164492A, HD164492B, HD164492C, HD164492D, and so on.
Spectral types are obtained from the Simbad database.
Diagnostic techniques employed to investigate the magnetic fields of stars with radiative cladding are primarily based on the Zeeman effect \citep{2021mfob.book.....H}.
The Zeeman effect causes atoms in a star's atmosphere to absorb energy at specific frequencies in the electromagnetic spectrum, generating absorption lines, which enables measurement of a star's magnetic field.
When subjected to a magnetic field, however, these spectral lines split into adjacent lines, producing polarized energy that depends on the original direction of the magnetic field.
As a result, the direction and intensity of a star's magnetic field can be determined by analyzing the spectral lines of the Zeeman effect \citep{2004IAUS..224..235W}.
Our dataset of stellar magnetic field observations consists of various observations of line Zeeman splitting and different observation methods, including Full Spectrum, Metal Line, H Balmer Line, Helium Line, and Least-Squares Deconvolution (LSD) procedures among others.
LSD was introduced by \cite{1997MNRAS.291..658D} to investigate the magnetic fields of late-active stars.
When recording observation methods, we include multiple observation and calculation methods under the label ``all'', while for a single method, we report data source information.
Column 10 of the catalogue represents the number of longitudinal magnetic field measurements employed in computing averaged quadratic effective magnetic field values.
As diverse studies contribute data to the catalogue, the source of longitudinal magnetic field employed for catalogue generation is presented in the last column, which provides useful information for thorough analyses.
\addtolength{\skip\footins}{-5pt}
\begin{table}
		\caption[]{The tiny part of the whole catalogue for 1784 magnetic stars which is given by Table 1 in the appendix.\footnotemark\label{Table1}}
	\setlength{\tabcolsep}{1pt}
	\small
	\begin{tabular}{ccccccccccc}
		\hline\noalign{\smallskip}
		Name& HD Number& RA& DEC& Sp type& $\left \langle B_{\rm e}\right \rangle$& $\left \langle\sigma_{\rm e}\right \rangle$& $\chi^2/n$& Method& Observation& References\\
		&&(J2000)&(J2000)&&($G$)&($G$)&&&count&\\
		\hline\noalign{\smallskip}
		HD 108 & 108 & 1.5141 & 63.6797 & O4-8f?p & 88 & 27 & 18.145  & all & 37 & \citealt{2017MNRAS.465.2432G} \\
		HD 166 & 166 & 1.6533 & +29.0215 & G0V & 2125 &  &  & metal & 1 & {\citealt{1983A&AS...53..119D}} \\
		4 Cet & 315 & 1.9338 & -2.5487 & ApSi & 1438 & 692 & 4.201  & metal & 4 & \citealt{2006MNRAS.372.1804K} \\
		...&...&...&...&...&...&...&...&...&...&...\\		
		\noalign{\smallskip}\hline
	\end{tabular}
\end{table}
\footnotetext{The Simbad Name, HD Number, ICRS(J2000) RA/DEC, and Stellar spectral type are shown in the first five columns, all which are obtained from the Simbad database. The sixth column gives the averaged quadratic effective magnetic field value $\left \langle B_{\rm e}\right \rangle$, and the seventh and eighth columns present the corresponding calculation error $\left \langle\sigma_{\rm e}\right \rangle$ and chi-square value $\chi^2/n$. The ninth column provides information on the spectral line type and calculation method of the Zeeman effect used for mean longitudinal magnetic field measurement, with the LSD procedure being the only calculation method. In the tenth column, we give the mean longitudinal magnetic field data count collected for each star, and in the last column, the relevant references are listed.}

\section{Statistic Analyse of Magnetic Stars}
\label{Sect3}
\cite{2003A&A...407..631B} employed the concept of averaged effective magnetic fields $\left \langle B_{\rm e}\right \rangle$ to establish two distinct correlations between the observed number of stars and their corresponding averaged quadratic effective magnetic field.
Here, for simplicity, $B=\left \langle B_{\rm e}\right \rangle$.
They demonstrated the relationship between the number distribution function $N(B)$ and its integral over $B$, with respect to the averaged quadratic effective magnetic field $B$.
In this context, $N(B)\rm{d}\emph{$B$}$ represents the averaged quadratic effective magnetic field $B$ across the range $[B,B+\rm{d}\emph{$B$}]$.
\cite{2003A&A...407..631B} defined the integrated distribution function as

\begin{equation} \label{eq4}
	N_{\rm Int}(B)=N_{\rm tot}-\int_{0}^{B}N(B')\rm{d}B',
\end{equation}

where the variable $N_{\rm tot}$ represents the total number of stars belonging to  a particular group.
\cite{2003A&A...407..631B} found that all the integrated distribution function $N_{\rm Int}(B)$ could be well approximated by the exponential function of the averaged quadratic effective magnetic field $B$

\begin{equation} \label{eq5}
	N_{\rm Int}(B)/N_{\rm tot}=\frac{a_1}{100\%}exp(-B/a_2),
\end{equation}

where coefficient $a_1$ (in \%) and $a_2$ (in G) depend on the class of magnetic stars.
In addition, \cite{2003A&A...407..631B} also investigated the number distribution function $N(B)$ of different type stars, where $N(B)$ can be given by

\begin{equation}\label{eq6}
	N(B)=-\frac{{\rm d}N_{\rm Int}}{{\rm d}B}.
\end{equation}

Then, $N(B)$ also can be an exponential function with the above analytic approximation by

\begin{equation}\label{eq7}
	N(B)=N_{\rm tot}\frac{a_1}{100\%}{a_2}^{-1}exp(-B/a_2).
\end{equation}

\begin{figure}[h]
	\centering
	\includegraphics[width=12.0cm, angle=0, scale=0.8]{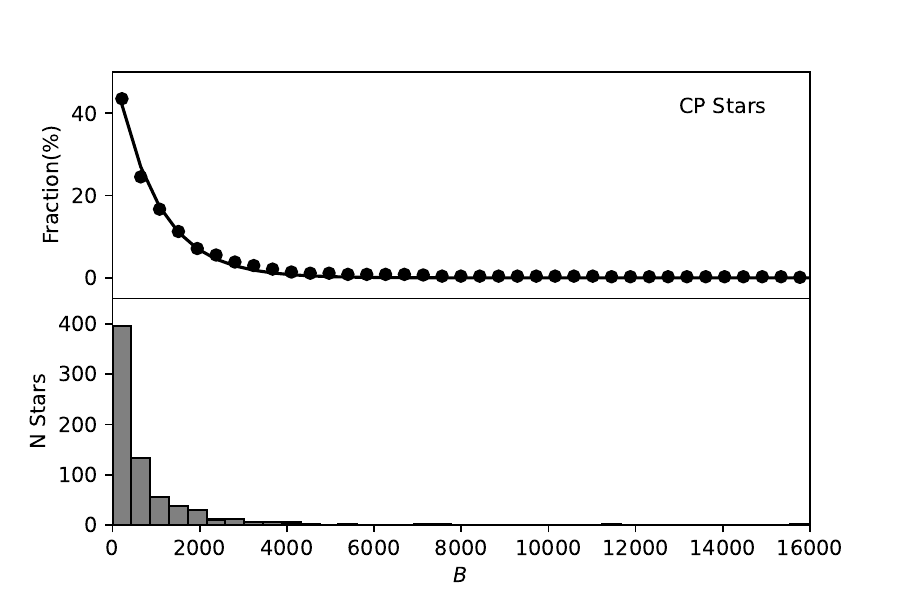}
	\caption{Integrated distribution function $N_{\rm Int}(B)$ in percent (upper panel), and the number distribution function $N(B)$ (lower panel) for 701 CP stars.\protect\footnotemark }
	\label{Fig3}
\end{figure}
\protect\footnotetext{The black dots represent the probability that upon investigating a new star of this type, its $\left \langle B_{\rm e}\right \rangle$ will be higher than the value of $B$. The solid line gives the best fit exponential function via the least square method. Similar with that in \cite{2003A&A...407..631B}, the range of the averaged quadratic effective magnetic field $\left \langle B_{\rm e}\right \rangle$ from zero to the maximum field into up to 80 equal-length bins and counted the number of stars in each bin.}

As an illustration, we discussed the relative integrated distribution function $N_{\rm Int}(B)$ in percentage and the number distribution function $N(B)$ concerning CP stars.
\cite{2003A&A...407..631B} focused on comprehensively exploring 352 CP stars, revealing the exponential-functional relationship between $N_{\rm Int}(B)$ and $N(B)$ with respect to the averaged quadratic effective magnetic field $B$ for these stars, where $a_1=97.2\%$ and $a_2=789.2$ $G$ for Eq. (\ref{eq5}).
By cross-matching 1784 magnetic stars with their corresponding CP stars in the present study, 307 CP stars were enumerated, where $N_{\rm Int}(B)$ and $N(B)$ are akin to those observed by
\newpage
\justify
\cite{2003A&A...407..631B} for 301 CP stars.
\cite{2009A&A...498..961R} curated an expanded catalogue of CP stars.
By cross-referencing 1784 magnetic stars with their compilation of 8205 CP stars, identified 701 CP stars in our catalogue.
Figure \ref{Fig3} showed the $N_{\rm Int}(B)$ and $N(B)$ of these 701 CP stars.
Through best-fit parameterization, we gave the exponential function, where $a_1=53\%$ and $a_2=972$ $G$ for Eq. (\ref{eq5}). The mean square error of the fitting curve is 0.25.
Compared with the exponential function for CP stars in \cite{2003A&A...407..631B}, the exponential function in the present paper has an lower exponent.
The main reasons are as follows: first of all, our sample is bigger.
In \cite{2003A&A...407..631B}, the maximum magnetic field is about 8000 $G$, while in the present paper, there are several stars whose magnetic field is higher than 8000 $G$, even the magnetic field of HD 215441 is up to 34543 $G$.
Secondly, in the present paper, we used the updated magnetic fields for some CP stars, for example, HD 358, HD 2453, HD 3980.
Nevertheless, despite these variances in exponents, the general distribution patterns of $N_{\rm Int}(B)$ and $N(B)$ remain similar.
As observations of CP stars continue to proliferate with increased observational accuracy, more precise distribution functions may be obtained.

	\begin{figure}[h]
	\centering
	\includegraphics[width=12.0cm, angle=0, scale=0.8]{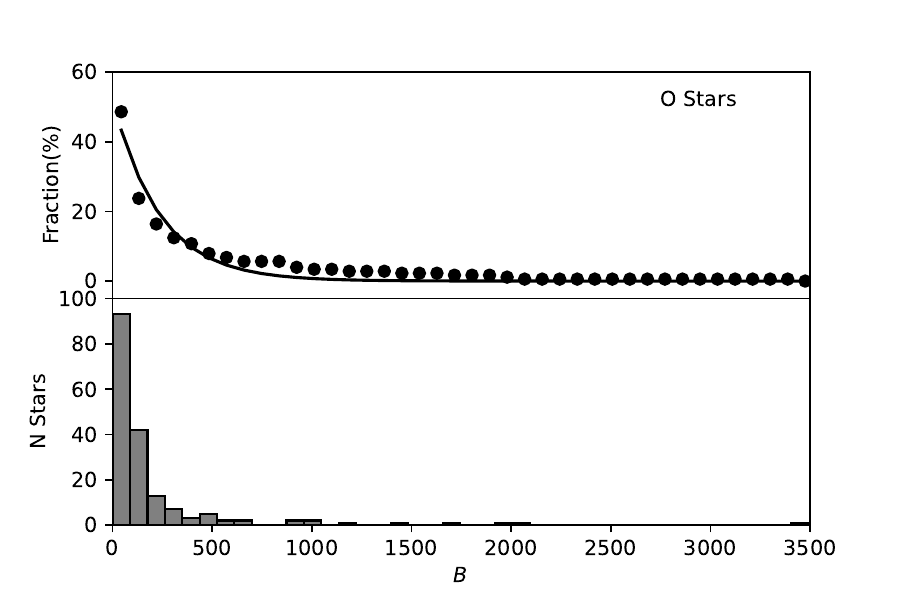}
	\caption{Similar to Figure \ref{Fig3}, but for 177 O-type magnetic stars.}
	\label{Fig4}
\end{figure}

	\begin{table}[h]
		\bc
		\caption[]{The table of optimal fitting parameters for  the exponential function (See Eq. (\ref{eq5})).\footnotemark  \label{Table2}}
		\setlength{\tabcolsep}{1pt}
		\small
		\begin{tabular}{ccccc}
			\hline\noalign{\smallskip}
			Spectral type& N& $a_1(\%)$& $a_2(G)$ & Mean Squared Error\\
			\hline\noalign{\smallskip}
			O Stars & 177 & 52 & 235 & 5.01\\
			B Stars & 551 & 40 & 678 & 0.70\\
			A Stars & 520 & 45 & 956 & 0.35\\
			F Stars & 91 & 36 & 813 & 4.09\\
			G Stars & 53 & 35 & 372 & 4.05\\
			K Stars & 61 & 33 & 463 & 8.52\\
			M Stars & 31 & 62 & 198 & 7.25\\
			\noalign{\smallskip}\hline
		\end{tabular}
		\ec
	\end{table}

	\footnotetext{In the first column, we highlight the spectral type of the stars. The second column provides the number of stars of each spectral type. The third and fourth columns show the best fitting parameters $a_1$ and $a_2$, respectively. The fifth column shows the mean square error. It's notable that coefficients $a_1(\%)$ and $a_2(G)$ are contingent on CP stars.}

\begin{figure}[t]
	\centering
	\includegraphics[width=12.0cm, angle=0, scale=0.8]{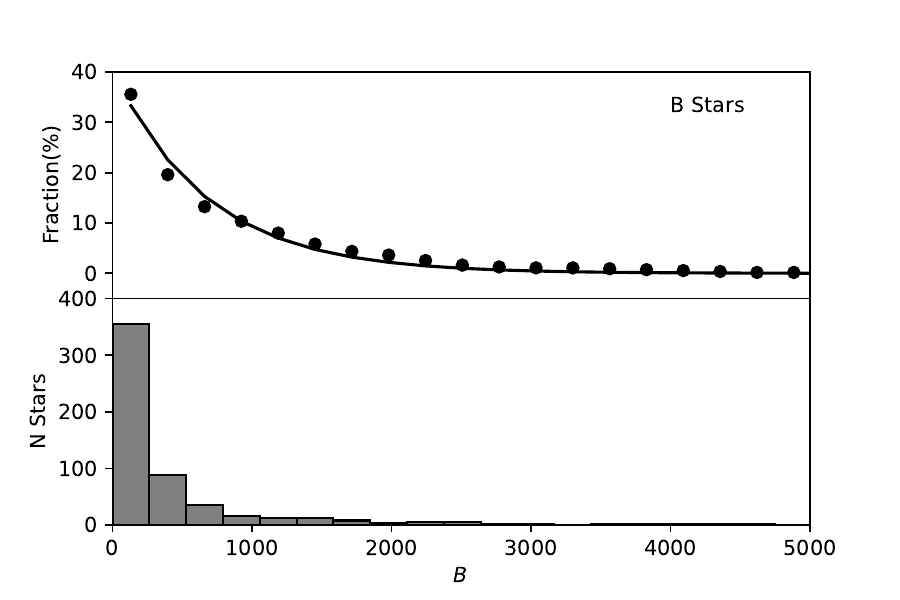}
	\caption{Similar to Figure \ref{Fig3}, but for 551 B-type magnetic stars.}
	\label{Fig5}
\end{figure}

\begin{figure}[t]
	\centering
	\includegraphics[width=12.0cm, angle=0, scale=0.8]{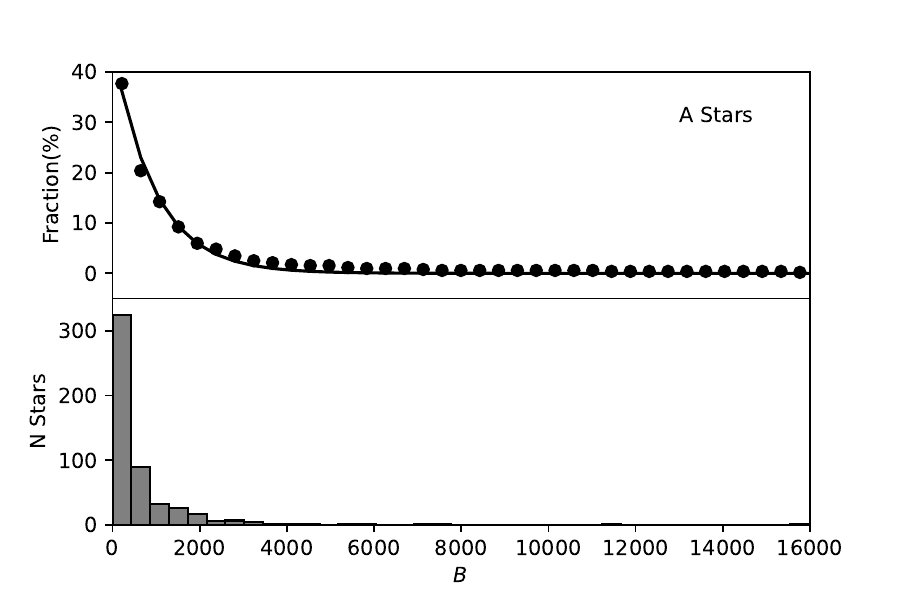}
	\caption{Similar to Figure \ref{Fig3}, but for 520 A-type magnetic stars.}
	\label{Fig6}
\end{figure}

\begin{figure}[t]
	\centering
	\includegraphics[width=12.0cm, angle=0, scale=0.8]{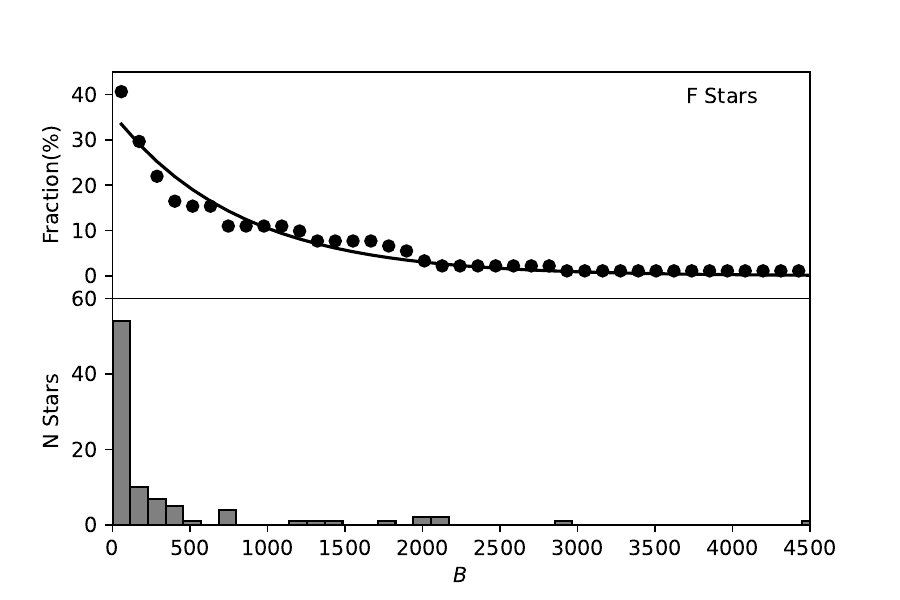}
	\caption{Similar to Figure \ref{Fig3}, but for 91 F-type magnetic stars.}
	\label{Fig7}
\end{figure}

\begin{figure}[t]
	\centering
	\includegraphics[width=12.0cm, angle=0, scale=0.8]{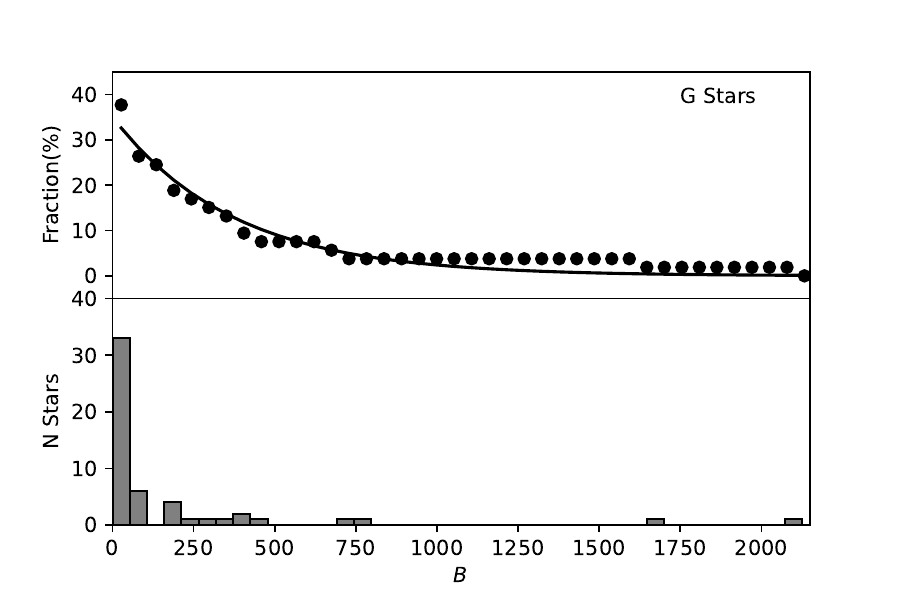}
	\caption{Similar to Figure \ref{Fig3}, but for 53 G-type magnetic stars.}
	\label{Fig8}
\end{figure}

\begin{figure}[t]
	\centering
	\includegraphics[width=12.0cm, angle=0, scale=0.8]{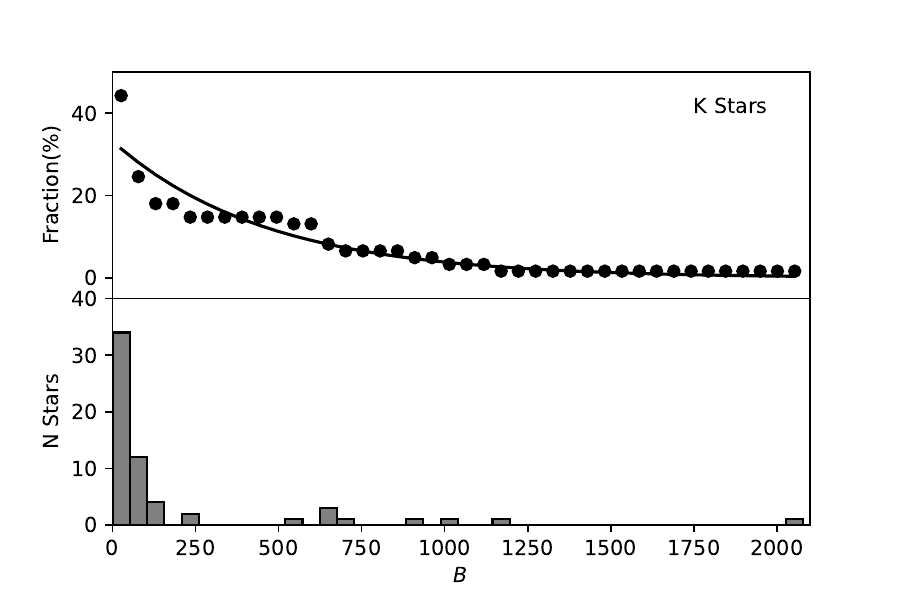}
	\caption{Similar to Figure \ref{Fig3}, but for 61 K-type magnetic stars.}
	\label{Fig9}
\end{figure}

\begin{figure}[t]
	\centering
	\includegraphics[width=12.0cm, angle=0, scale=0.8]{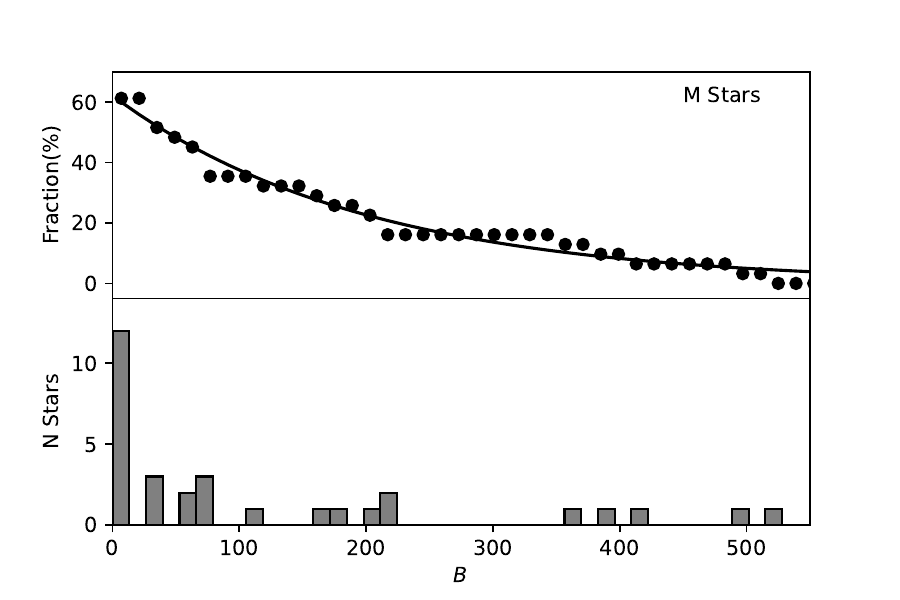}
	\caption{Similar to Figure \ref{Fig3}, but for 31 K-type magnetic stars.}
	\label{Fig10}
\end{figure}

In this study, we presented spectral types for all magnetic stars.
Figures \ref{Fig4}-\ref{Fig10} exhibit the relative integrated distribution function $N_{\rm Int}(B)$ (upper panels) and number distribution function (lower panels) for O, B, A, F, G, K, and M type stars, respectively.
Analogous to CP stars, both of these distributions can be suitably described by an exponential function of the averaged quadratic effective magnetic field $B$.
The parameters $a_1$ and $a_2$, along with the mean square error, obtained from the best-fit analysis, are listed in Table \ref{Table2}.
While the origin of stellar magnetic fields remains controversial, it is well established that for lower mass or cooler stars, cyclonic turbulence and rotational shearing in the convective zone account for the magnetic field \citep{1995ApJ...440..415P}.
For massive stars, the magnetic field may originate from fossil magnetic field or binary merger \citep{2019Natur.574..211S}.
Intriguingly, our results show that the integrated distribution function and number distribution function of both low-mass and high-mass stars converge towards a similar exponential function of $B$.
This may suggest that the magnetic origin of all stars is alike.
Of course, these similar distributions could possibly be an artifact of observational bias.
It is beyond the scope of this paper to find the origin of magnetic stars.

\begin{figure}
	\centering
	\includegraphics[width=12.0cm, angle=0, scale=0.8]{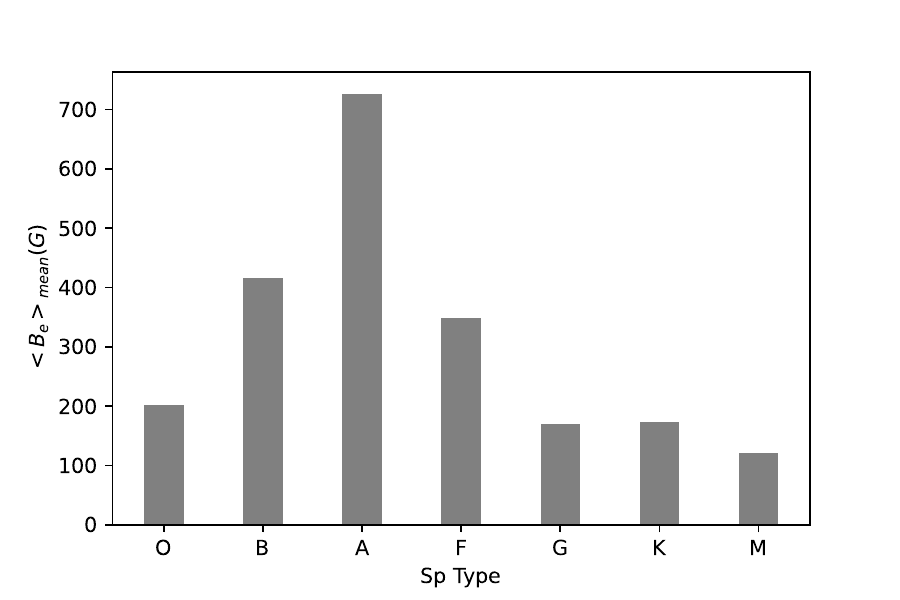}
	\caption{ The averaged magnetic field calculated by Eq. (\ref{eq8}) for every spectral type magnetic stars.}
	\label{Fig11}
\end{figure}

Furthermore, the mean magnetic field strength for each spectral type averaged quadratic effective magnetic fields can be calculated as

	\begin{equation} \label{eq8}
		\left \langle B_{\rm e}\right \rangle_{\rm mean} 	
		=\frac{1}{n}\sum_{i=1}^n\left \langle B_{\rm e}\right \rangle_{i}.
	\end{equation}
Figure \ref{Fig11} illustrates the mean magnetic field strength $\left \langle B_{\rm e}\right \rangle_{\rm mean}$ for a range of spectral types, including O,
\newpage
\justify
B, A, F, G, K, and M stars.
Evidently, the average magnetic fields of A and B type stars are stronger among all type of stars.
This observation is noteworthy, usually, the stronger the magnetic field is, the more sign-
\newpage
\justify
ificant the Zeeman Effect is.
Consequently, the detectability of magnetic fields in A and B stars is higher relative to other stars, a trend that is also substantiated by Figure \ref{Fig2}.

\section{Conclusions}
\label{Sect4}
This study presents a comprehensive catalogue of 1784 magnetic stars, including their names, HD numbers, coordinates, spectral types, averaged quadratic effective magnetic fields, corresponding calculation errors, chi-square values, calculation methods, observational counts for each magnetic star, and references.
These magnetic stars comprise 177 O-type stars, 551 B-type stars, 520 A-type stars, 91 F-type stars, 53 G-type stars, 61 K-type stars, 31 M-type stars, along with an additional 300 stars whose spectral classification remains unidentified.
Our analysis encompasses the statistical characteristics of these magnetic stars.
We discovered that, akin to CP stars observed by \cite{2003A&A...407..631B}, both $N_{\rm Int}(B)$ and $N(B)$ for magnetic stars possessing the same spectral types can be suitably approximated via exponential functions of the averaged quadratic effective magnetic field $B$.
Additionally, we determined the mean magnetic field for each spectral type, revealing that A and B type stars exhibit the strongest mean magnetic fields.
Consistent
\newpage
\justify
with observed magnetic stars, this outcome suggests that magnetic fields in A and B type stars are more readily detectable relative to those of other spectral types.

\normalem
\begin{acknowledgements}
This work received the generous support of the National Natural Science Foundation of China under grants U2031204, 12163005, and 12288102; the Natural Science Foundation of Xinjiang Nos. 2021D01C075, 2022D01D85, 2022TSYCLJ0006 and 2022D01E86; the science research grants from the China Manned Space Project with Nos.CMS-CSST-2021-A10 and CMS-CSST-2021-A08; the Chinese Academy of Sciences (CAS) “Light of West China" Program, No. 2022-XBQNXZ-013.

\end{acknowledgements}

\bibliographystyle{raa}
\bibliography{RAA-2023-0199BibTeX} 

\begin{thebibliography}{242}
\providecommand\natexlab[1]{#1}
\providecommand\JournalTitle[1]{#1}

\bibitem[{Adam}(1965)]{1965Obs....85..204A}
{Adam}, M.~G. 1965, The Observatory, 85, 204

\bibitem[{Albrecht} {et~al.}(1977)]{1977A&A....58...93A}
{Albrecht}, R., {Jenkner}, H., {Weiss}, W.~W., \& {Wood}, H.~J. 1977, \aap, 58,
  93

\bibitem[{Angel} {et~al.}(1973)]{1973ApJ...184L..79A}
{Angel}, J.~R.~P., {McGraw}, J.~T., \& {Stockman}, H.~S., J. 1973, \apjl, 184,
  L79

\bibitem[{Auri{\`e}re} {et~al.}(2007)]{2007A&A...475.1053A}
{Auri{\`e}re}, M., {Wade}, G.~A., {Silvester}, J., {et~al.} 2007, \aap, 475,
  1053

\bibitem[{Aznar Cuadrado} {et~al.}(2004)]{2004A&A...423.1081A}
{Aznar Cuadrado}, R., {Jordan}, S., {Napiwotzki}, R., {et~al.} 2004, \aap, 423,
  1081

\bibitem[{Aznar Cuadrado} {et~al.}(2005)]{2005ASPC..334..159A}
{Aznar Cuadrado}, R., {Jordan}, S., {Napiwotzki}, R., {et~al.} 2005, in
  Astronomical Society of the Pacific Conference Series, Vol. 334, 14th
  European Workshop on White Dwarfs, ed. D.~{Koester} \& S.~{Moehler}, 159

\bibitem[{Babcock}(1954)]{1954ApJ...120...66B}
{Babcock}, H.~W. 1954, \apj, 120, 66

\bibitem[{Babcock}(1956)]{1956ApJ...124..489B}
{Babcock}, H.~W. 1956, \apj, 124, 489

\bibitem[{Babcock}(1958)]{1958ApJS....3..141B}
{Babcock}, H.~W. 1958, \apjs, 3, 141

\bibitem[{Babcock}(1960)]{1960ApJ...132..521B}
{Babcock}, H.~W. 1960, \apj, 132, 521

\bibitem[{Babcock} \& {Burd}(1952)]{1952ApJ...116....8B}
{Babcock}, H.~W., \& {Burd}, S. 1952, \apj, 116, 8

\bibitem[{Babel} \& {North}(1997)]{1997A&A...325..195B}
{Babel}, J., \& {North}, P. 1997, \aap, 325, 195

\bibitem[{Bagnulo} {et~al.}(2015)]{2015A&A...583A.115B}
{Bagnulo}, S., {Fossati}, L., {Landstreet}, J.~D., \& {Izzo}, C. 2015, \aap,
  583, A115

\bibitem[{Bagnulo} {et~al.}(2004)]{2004A&A...416.1149B}
{Bagnulo}, S., {Hensberge}, H., {Landstreet}, J.~D., {Szeifert}, T., \& {Wade},
  G.~A. 2004, \aap, 416, 1149

\bibitem[{Bagnulo} {et~al.}(2012)]{2012A&A...538A.129B}
{Bagnulo}, S., {Landstreet}, J.~D., {Fossati}, L., \& {Kochukhov}, O. 2012,
  \aap, 538, A129

\bibitem[{Bagnulo} {et~al.}(2003)]{2003A&A...403..645B}
{Bagnulo}, S., {Landstreet}, J.~D., {Lo Curto}, G., {Szeifert}, T., \& {Wade},
  G.~A. 2003, \aap, 403, 645

\bibitem[{Bagnulo} {et~al.}(2006)]{2006A&A...450..777B}
{Bagnulo}, S., {Landstreet}, J.~D., {Mason}, E., {et~al.} 2006, \aap, 450, 777

\bibitem[{Bagnulo} {et~al.}(2005)]{2005ASPC..343..369B}
{Bagnulo}, S., {Mason}, E., {Szeifert}, T., {et~al.} 2005, in Astronomical
  Society of the Pacific Conference Series, Vol. 343, Astronomical Polarimetry:
  Current Status and Future Directions, ed. A.~{Adamson}, C.~{Aspin},
  C.~{Davis}, \& T.~{Fujiyoshi}, 369

\bibitem[{Bagnulo} {et~al.}(2002)]{2002A&A...389..191B}
{Bagnulo}, S., {Szeifert}, T., {Wade}, G.~A., {Landstreet}, J.~D., \& {Mathys},
  G. 2002, \aap, 389, 191

\bibitem[{Bagnulo} {et~al.}(2017)]{2017A&A...601A.136B}
{Bagnulo}, S., {Naz{\'e}}, Y., {Howarth}, I.~D., {et~al.} 2017, \aap, 601, A136

\bibitem[{Bagnulo} {et~al.}(2020)]{2020A&A...635A.163B}
{Bagnulo}, S., {Wade}, G.~A., {Naz{\'e}}, Y., {et~al.} 2020, \aap, 635, A163

\bibitem[{Barker} {et~al.}(1985)]{1985ApJ...288..741B}
{Barker}, P.~K., {Landstreet}, J.~D., {Marlborough}, J.~M., \& {Thompson},
  I.~B. 1985, \apj, 288, 741

\bibitem[{Bedford} {et~al.}(1995)]{1995A&A...293..377B}
{Bedford}, D.~K., {Chaplin}, W.~J., {Davies}, A.~R., {et~al.} 1995, \aap, 293,
  377

\bibitem[{Bedford} {et~al.}(1994)]{1994MNRAS.269..639B}
{Bedford}, D.~K., {Chaplin}, W.~J., {Innis}, J.~L., {et~al.} 1994, \mnras, 269,
  639

\bibitem[{Boesgaard}(1974)]{1974ApJ...188..567B}
{Boesgaard}, A.~M. 1974, \apj, 188, 567

\bibitem[{Boesgaard} {et~al.}(1975)]{1975PASP...87..353B}
{Boesgaard}, A.~M., {Chesley}, D., \& {Preston}, G.~W. 1975, \pasp, 87, 353

\bibitem[{Bohlender}(1989)]{1989A&A...220..215B}
{Bohlender}, D.~A. 1989, \aap, 220, 215

\bibitem[{Bohlender} {et~al.}(1987)]{1987ApJ...323..325B}
{Bohlender}, D.~A., {Brown}, D.~N., {Landstreet}, J.~D., \& {Thompson}, I.~B.
  1987, \apj, 323, 325

\bibitem[{Bohlender} \& {Landstreet}(1990{\natexlab{a}})]{1990MNRAS.247..606B}
{Bohlender}, D.~A., \& {Landstreet}, J.~D. 1990{\natexlab{a}}, \mnras, 247, 606

\bibitem[{Bohlender} \& {Landstreet}(1990{\natexlab{b}})]{1990ApJ...358L..25B}
{Bohlender}, D.~A., \& {Landstreet}, J.~D. 1990{\natexlab{b}}, \apjl, 358, L25

\bibitem[{Bohlender} {et~al.}(1993)]{1993A&A...269..355B}
{Bohlender}, D.~A., {Landstreet}, J.~D., \& {Thompson}, I.~B. 1993, \aap, 269,
  355

\bibitem[{Bonsack}(1976)]{1976ApJ...209..160B}
{Bonsack}, W.~K. 1976, \apj, 209, 160

\bibitem[{Bonsack}(1977)]{1977A&A....59..195B}
{Bonsack}, W.~K. 1977, \aap, 59, 195

\bibitem[{Bonsack}(1981)]{1981PASP...93..756B}
{Bonsack}, W.~K. 1981, \pasp, 93, 756

\bibitem[{Bonsack} \& {Pilachowski}(1974)]{1974ApJ...190..327B}
{Bonsack}, W.~K., \& {Pilachowski}, C.~A. 1974, \apj, 190, 327

\bibitem[{Bonsack} {et~al.}(1974)]{1974ApJ...187..265B}
{Bonsack}, W.~K., {Pilachowski}, C.~A., \& {Wolff}, S.~C. 1974, \apj, 187, 265

\bibitem[{Bonsack} \& {Simon}(1983)]{1983IAUS..102...35B}
{Bonsack}, W.~K., \& {Simon}, T. 1983, in Solar and Stellar Magnetic Fields:
  Origins and Coronal Effects, ed. J.~O. {Stenflo}, Vol. 102, 35

\bibitem[{Borra}(1975)]{1975ApJ...202..741B}
{Borra}, E.~F. 1975, \apj, 202, 741

\bibitem[{Borra}(1980)]{1980ApJ...235..911B}
{Borra}, E.~F. 1980, \apj, 235, 911

\bibitem[{Borra}(1981)]{1981ApJ...249L..39B}
{Borra}, E.~F. 1981, \apjl, 249, L39

\bibitem[{Borra} {et~al.}(1984)]{1984ApJ...284..211B}
{Borra}, E.~F., {Edwards}, G., \& {Mayor}, M. 1984, \apj, 284, 211

\bibitem[{Borra} {et~al.}(1981)]{1981ApJ...247..569B}
{Borra}, E.~F., {Fletcher}, J.~M., \& {Poeckert}, R. 1981, \apj, 247, 569

\bibitem[{Borra} \& {Landstreet}(1973)]{1973ApJ...185L.139B}
{Borra}, E.~F., \& {Landstreet}, J.~D. 1973, \apjl, 185, L139

\bibitem[{Borra} \& {Landstreet}(1975)]{1975PASP...87..961B}
{Borra}, E.~F., \& {Landstreet}, J.~D. 1975, \pasp, 87, 961

\bibitem[{Borra} \& {Landstreet}(1977)]{1977ApJ...212..141B}
{Borra}, E.~F., \& {Landstreet}, J.~D. 1977, \apj, 212, 141

\bibitem[{Borra} \& {Landstreet}(1978)]{1978ApJ...222..226B}
{Borra}, E.~F., \& {Landstreet}, J.~D. 1978, \apj, 222, 226

\bibitem[{Borra} \& {Landstreet}(1979)]{1979ApJ...228..809B}
{Borra}, E.~F., \& {Landstreet}, J.~D. 1979, \apj, 228, 809

\bibitem[{Borra} \& {Landstreet}(1980)]{1980ApJS...42..421B}
{Borra}, E.~F., \& {Landstreet}, J.~D. 1980, \apjs, 42, 421

\bibitem[{Borra} {et~al.}(1983)]{1983ApJS...53..151B}
{Borra}, E.~F., {Landstreet}, J.~D., \& {Thompson}, I. 1983, \apjs, 53, 151

\bibitem[{Borra} {et~al.}(1973)]{1973ApJ...185L.145B}
{Borra}, E.~F., {Landstreet}, J.~D., \& {Vaughan}, A.~H., J. 1973, \apjl, 185,
  L145

\bibitem[{Borra} \& {Vaughan}(1977)]{1977ApJ...216..462B}
{Borra}, E.~F., \& {Vaughan}, A.~H. 1977, \apj, 216, 462

\bibitem[{Briquet} {et~al.}(2007)]{2007AN....328...41B}
{Briquet}, M., {Hubrig}, S., {Sch{\"o}ller}, M., \& {De Cat}, P. 2007,
  Astronomische Nachrichten, 328, 41

\bibitem[{Brown} \& {Landstreet}(1981)]{1981ApJ...246..899B}
{Brown}, D.~N., \& {Landstreet}, J.~D. 1981, \apj, 246, 899

\bibitem[{Brown} {et~al.}(1985)]{1985AJ.....90.1354B}
{Brown}, D.~N., {Shore}, S.~N., \& {Sonneborn}, G. 1985, \aj, 90, 1354

\bibitem[{Bychkov} {et~al.}(2003)]{2003A&A...407..631B}
{Bychkov}, V.~D., {Bychkova}, L.~V., \& {Madej}, J. 2003, \aap, 407, 631

\bibitem[{Bychkov} {et~al.}(2006)]{2006MNRAS.365..585B}
{Bychkov}, V.~D., {Bychkova}, L.~V., \& {Madej}, J. 2006, \mnras, 365, 585

\bibitem[{Bychkov} {et~al.}(2009)]{2009MNRAS.394.1338B}
{Bychkov}, V.~D., {Bychkova}, L.~V., \& {Madej}, J. 2009, \mnras, 394, 1338

\bibitem[{Castro} {et~al.}(2017)]{2017A&A...597L...6C}
{Castro}, N., {Fossati}, L., {Hubrig}, S., {et~al.} 2017, \aap, 597, L6

\bibitem[{Catala} {et~al.}(2007)]{2007A&A...462..293C}
{Catala}, C., {Alecian}, E., {Donati}, J.~F., {et~al.} 2007, \aap, 462, 293

\bibitem[{Chadid} {et~al.}(2004)]{2004A&A...413.1087C}
{Chadid}, M., {Wade}, G.~A., {Shorlin}, S.~L.~S., \& {Landstreet}, J.~D. 2004,
  \aap, 413, 1087

\bibitem[{Charbonneau}(2010)]{2010LRSP....7....3C}
{Charbonneau}, P. 2010, Living Reviews in Solar Physics, 7, 3

\bibitem[{Conti}(1969)]{1969ApJ...156..661C}
{Conti}, P.~S. 1969, \apj, 156, 661

\bibitem[{Conti}(1970{\natexlab{a}})]{1970ApJ...160.1077C}
{Conti}, P.~S. 1970{\natexlab{a}}, \apj, 160, 1077

\bibitem[{Conti}(1970{\natexlab{b}})]{1970ApJ...159..723C}
{Conti}, P.~S. 1970{\natexlab{b}}, \apj, 159, 723

\bibitem[{de Jong} {et~al.}(2001)]{2001A&A...368..601D}
{de Jong}, J.~A., {Henrichs}, H.~F., {Kaper}, L., {et~al.} 2001, \aap, 368, 601

\bibitem[{Didelon}(1983)]{1983A&AS...53..119D}
{Didelon}, P. 1983, \aaps, 53, 119

\bibitem[{Donati} \& {Landstreet}(2009)]{2009ARA&A..47..333D}
{Donati}, J.~F., \& {Landstreet}, J.~D. 2009, \araa, 47, 333

\bibitem[{Donati} {et~al.}(1997)]{1997MNRAS.291..658D}
{Donati}, J.~F., {Semel}, M., {Carter}, B.~D., {Rees}, D.~E., \& {Collier
  Cameron}, A. 1997, \mnras, 291, 658

\bibitem[{Donati} {et~al.}(2006)]{2006MNRAS.370..629D}
{Donati}, J.~F., {Howarth}, I.~D., {Jardine}, M.~M., {et~al.} 2006, \mnras,
  370, 629

\bibitem[{Elkin}(1996)]{1996A&A...312L...5E}
{Elkin}, V.~G. 1996, \aap, 312, L5

\bibitem[{Elkin}(1998)]{1998CoSka..27..452E}
{Elkin}, V.~G. 1998, Contributions of the Astronomical Observatory Skalnate
  Pleso, 27, 452

\bibitem[{El'kin}(1999)]{1999AstL...25..809E}
{El'kin}, V.~G. 1999, Astronomy Letters, 25, 809

\bibitem[{El'Kin} {et~al.}(2002)]{2002AstL...28..169E}
{El'Kin}, V.~G., {Kudryavtsev}, D.~O., \& {Romanyuk}, I.~I. 2002, Astronomy
  Letters, 28, 169

\bibitem[{El'Kin} {et~al.}(2003)]{2003AstL...29..400E}
{El'Kin}, V.~G., {Kudryavtsev}, D.~O., \& {Romanyuk}, I.~I. 2003, Astronomy
  Letters, 29, 400

\bibitem[{Fabrika} {et~al.}(2003)]{2003AstL...29..737F}
{Fabrika}, S.~N., {Valyavin}, G.~G., \& {Burlakova}, T.~E. 2003, Astronomy
  Letters, 29, 737

\bibitem[{Fossati} {et~al.}(2015{\natexlab{a}})]{2015A&A...582A..45F}
{Fossati}, L., {Castro}, N., {Sch{\"o}ller}, M., {et~al.} 2015{\natexlab{a}},
  \aap, 582, A45

\bibitem[{Fossati} {et~al.}(2015{\natexlab{b}})]{2015A&A...574A..20F}
{Fossati}, L., {Castro}, N., {Morel}, T., {et~al.} 2015{\natexlab{b}}, \aap,
  574, A20

\bibitem[{Gerth} {et~al.}(1999)]{1999A&A...351..133G}
{Gerth}, E., {Glagolevskij}, Y.~V., {Hildebrandt}, G., {Lehmann}, H., \&
  {Scholz}, G. 1999, \aap, 351, 133

\bibitem[{Gerth} {et~al.}(1991)]{1991AN....312..107G}
{Gerth}, E., {Scholz}, G., {Glagolevskii}, I.~V., \& {Romaniuk}, I.~I. 1991,
  Astronomische Nachrichten, 312, 107

\bibitem[{Glagolevskii} {et~al.}(1977)]{1977SvAL....3..273G}
{Glagolevskii}, I.~V., {Kozlova}, K.~I., {Kopylov}, I.~M., {et~al.} 1977,
  Soviet Astronomy Letters, 3, 273

\bibitem[{Glagolevskij} {et~al.}(1982)]{1982SvAL....8...12G}
{Glagolevskij}, Y.~V., {Bychkov}, V.~D., {Iliev}, I.~K., {Romanyuk}, I.~I., \&
  {Chunakova}, N.~M. 1982, Soviet Astronomy Letters, 8, 12

\bibitem[{Glagolevskij} {et~al.}(1998)]{1998CoSka..27..458G}
{Glagolevskij}, Y.~V., {Gerth}, E., {Hildebrandt}, G., {Lehmann}, H., \&
  {Scholz}, G. 1998, Contributions of the Astronomical Observatory Skalnate
  Pleso, 27, 458

\bibitem[{Gollnow}(1962)]{1962PASP...74..163G}
{Gollnow}, H. 1962, \pasp, 74, 163

\bibitem[{Gollnow}(1971)]{1971Obs....91...37G}
{Gollnow}, H. 1971, The Observatory, 91, 37

\bibitem[{Gonz{\'a}lez} {et~al.}(2017)]{2017MNRAS.467..437G}
{Gonz{\'a}lez}, J.~F., {Hubrig}, S., {Przybilla}, N., {et~al.} 2017, \mnras,
  467, 437

\bibitem[{Grunhut} {et~al.}(2017)]{2017MNRAS.465.2432G}
{Grunhut}, J.~H., {Wade}, G.~A., {Neiner}, C., {et~al.} 2017, \mnras, 465, 2432

\bibitem[{Hale}(1908)]{1908ApJ....28..315H}
{Hale}, G.~E. 1908, \apj, 28, 315

\bibitem[{Hildebrandt} {et~al.}(1973)]{1973AN....294..175H}
{Hildebrandt}, G., {Nikolow}, A.~S., {Scholz}, G., \& {Sch{\"o}neich}, W. 1973,
  Astronomische Nachrichten, 294, 175

\bibitem[{Hill} \& {Blake}(1996)]{1996MNRAS.278..183H}
{Hill}, G.~M., \& {Blake}, C.~C. 1996, \mnras, 278, 183

\bibitem[{Hockey}(1969)]{1969MNRAS.142..543H}
{Hockey}, M.~S. 1969, \mnras, 142, 543

\bibitem[{Hockey}(1971)]{1971MNRAS.152...97H}
{Hockey}, M.~S. 1971, \mnras, 152, 97

\bibitem[{Hubrig} {et~al.}(2006{\natexlab{a}})]{2006MNRAS.369L..61H}
{Hubrig}, S., {Briquet}, M., {Sch{\"o}ller}, M., {et~al.} 2006{\natexlab{a}},
  \mnras, 369, L61

\bibitem[{Hubrig} {et~al.}(2007{\natexlab{a}})]{2007ASPC..361..434H}
{Hubrig}, S., {Briquet}, M., {Sch{\"o}ller}, M., {et~al.} 2007{\natexlab{a}},
  in Astronomical Society of the Pacific Conference Series, Vol. 361, Active
  OB-Stars: Laboratories for Stellare and Circumstellar Physics, ed. A.~T.
  {Okazaki}, S.~P. {Owocki}, \& S.~{Stefl}, 434

\bibitem[{Hubrig} {et~al.}(2004{\natexlab{a}})]{2004A&A...415..661H}
{Hubrig}, S., {Kurtz}, D.~W., {Bagnulo}, S., {et~al.} 2004{\natexlab{a}}, \aap,
  415, 661

\bibitem[{Hubrig} {et~al.}(2006{\natexlab{b}})]{2006AN....327..289H}
{Hubrig}, S., {North}, P., {Sch{\"o}ller}, M., \& {Mathys}, G.
  2006{\natexlab{b}}, Astronomische Nachrichten, 327, 289

\bibitem[{Hubrig} {et~al.}(1994)]{1994A&A...291..890H}
{Hubrig}, S., {Plachinda}, S.~I., {Hunsch}, M., \& {Schroder}, K.~P. 1994,
  \aap, 291, 890

\bibitem[{Hubrig} {et~al.}(2007{\natexlab{b}})]{2007A&A...463.1039H}
{Hubrig}, S., {Pogodin}, M.~A., {Yudin}, R.~V., {Sch{\"o}ller}, M., \&
  {Schnerr}, R.~S. 2007{\natexlab{b}}, \aap, 463, 1039

\bibitem[{Hubrig} \& {Sch{\"o}ller}(2021)]{2021mfob.book.....H}
{Hubrig}, S., \& {Sch{\"o}ller}, M. 2021, {Magnetic Fields in O, B, and A
  Stars}

\bibitem[{Hubrig} {et~al.}(2020)]{2020MNRAS.499L.116H}
{Hubrig}, S., {Sch{\"o}ller}, M., {Cikota}, A., \& {J{\"a}rvinen}, S.~P. 2020,
  \mnras, 499, L116

\bibitem[{Hubrig} {et~al.}(2008)]{2008A&A...490..793H}
{Hubrig}, S., {Sch{\"o}ller}, M., {Schnerr}, R.~S., {et~al.} 2008, \aap, 490,
  793

\bibitem[{Hubrig} {et~al.}(2004{\natexlab{b}})]{2004A&A...428L...1H}
{Hubrig}, S., {Sch{\"o}ller}, M., \& {Yudin}, R.~V. 2004{\natexlab{b}}, \aap,
  428, L1

\bibitem[{Hubrig} {et~al.}(2004{\natexlab{c}})]{2004A&A...415..685H}
{Hubrig}, S., {Szeifert}, T., {Sch{\"o}ller}, M., {Mathys}, G., \& {Kurtz},
  D.~W. 2004{\natexlab{c}}, \aap, 415, 685

\bibitem[{Hubrig} {et~al.}(2006{\natexlab{c}})]{2006A&A...446.1089H}
{Hubrig}, S., {Yudin}, R.~V., {Sch{\"o}ller}, M., \& {Pogodin}, M.~A.
  2006{\natexlab{c}}, \aap, 446, 1089

\bibitem[{Hubrig} {et~al.}(2005)]{2005A&A...440L..37H}
{Hubrig}, S., {Nesvacil}, N., {Sch{\"o}ller}, M., {et~al.} 2005, \aap, 440, L37

\bibitem[{Hubrig} {et~al.}(2013)]{2013A&A...551A..33H}
{Hubrig}, S., {Sch{\"o}ller}, M., {Ilyin}, I., {et~al.} 2013, \aap, 551, A33

\bibitem[{Hubrig} {et~al.}(2014)]{2014A&A...564L..10H}
{Hubrig}, S., {Fossati}, L., {Carroll}, T.~A., {et~al.} 2014, \aap, 564, L10

\bibitem[{Hubrig} {et~al.}(2015)]{2015A&A...578L...3H}
{Hubrig}, S., {Sch{\"o}ller}, M., {Fossati}, L., {et~al.} 2015, \aap, 578, L3

\bibitem[{J{\"a}rvinen} {et~al.}(2019)]{2019MNRAS.486.5499J}
{J{\"a}rvinen}, S.~P., {Carroll}, T.~A., {Hubrig}, S., {et~al.} 2019, \mnras,
  486, 5499

\bibitem[{J{\"a}rvinen} {et~al.}(2018)]{2018MNRAS.481.5163J}
{J{\"a}rvinen}, S.~P., {Hubrig}, S., {Scholz}, R.~D., {et~al.} 2018, \mnras,
  481, 5163

\bibitem[{Johnstone} \& {Penston}(1986)]{1986MNRAS.219..927J}
{Johnstone}, R.~M., \& {Penston}, M.~V. 1986, \mnras, 219, 927

\bibitem[{Johnstone} \& {Penston}(1987)]{1987MNRAS.227..797J}
{Johnstone}, R.~M., \& {Penston}, M.~V. 1987, \mnras, 227, 797

\bibitem[{Jones} \& {Wolff}(1974)]{1974PASP...86...67J}
{Jones}, T.~J., \& {Wolff}, S.~C. 1974, \pasp, 86, 67

\bibitem[{Jones} {et~al.}(1974)]{1974ApJ...190..579J}
{Jones}, T.~J., {Wolff}, S.~C., \& {Bonsack}, W.~K. 1974, \apj, 190, 579

\bibitem[{Jordan} {et~al.}(2012)]{2012A&A...542A..64J}
{Jordan}, S., {Bagnulo}, S., {Werner}, K., \& {O'Toole}, S.~J. 2012, \aap, 542,
  A64

\bibitem[{Keszthelyi}(2023)]{2023Galax..11...40K}
{Keszthelyi}, Z. 2023, Galaxies, 11, 40

\bibitem[{Kim} {et~al.}(2007)]{2007PASP..119.1052K}
{Kim}, K.-M., {Han}, I., {Valyavin}, G.~G., {et~al.} 2007, \pasp, 119, 1052

\bibitem[{Kippenhahn} \& {Weigert}(1990)]{1990sse..book.....K}
{Kippenhahn}, R., \& {Weigert}, A. 1990, {Stellar Structure and Evolution}

\bibitem[{Kochukhov} \& {Bagnulo}(2006)]{2006A&A...450..763K}
{Kochukhov}, O., \& {Bagnulo}, S. 2006, \aap, 450, 763

\bibitem[{Kochukhov} \& {Wade}(2007)]{2007A&A...467..679K}
{Kochukhov}, O., \& {Wade}, G.~A. 2007, \aap, 467, 679

\bibitem[{Kodaira} \& {Unno}(1969)]{1969ApJ...157..769K}
{Kodaira}, K., \& {Unno}, W. 1969, \apj, 157, 769

\bibitem[{Kudryavtsev} {et~al.}(2006)]{2006MNRAS.372.1804K}
{Kudryavtsev}, D.~O., {Romanyuk}, I.~I., {Elkin}, V.~G., \& {Paunzen}, E. 2006,
  \mnras, 372, 1804

\bibitem[{Kuvshinov} {et~al.}(1976)]{1976AN....297..181K}
{Kuvshinov}, V.~M., {Hildebrandt}, G., \& {Schoeneich}, W. 1976, Astronomische
  Nachrichten, 297, 181

\bibitem[{Landstreet}(1982)]{1982ApJ...258..639L}
{Landstreet}, J.~D. 1982, \apj, 258, 639

\bibitem[{Landstreet}(1990)]{1990ApJ...352L...5L}
{Landstreet}, J.~D. 1990, \apjl, 352, L5

\bibitem[{Landstreet}(1992)]{1992A&ARv...4...35L}
{Landstreet}, J.~D. 1992, \aapr, 4, 35

\bibitem[{Landstreet} \& {Borra}(1977)]{1977ApJ...212L..43L}
{Landstreet}, J.~D., \& {Borra}, E.~F. 1977, \apjl, 212, L43

\bibitem[{Landstreet} \& {Borra}(1978)]{1978ApJ...224L...5L}
{Landstreet}, J.~D., \& {Borra}, E.~F. 1978, \apjl, 224, L5

\bibitem[{Landstreet} {et~al.}(1975)]{1975ApJ...201..624L}
{Landstreet}, J.~D., {Borra}, E.~F., {Angel}, J.~R.~P., \& {Illing}, R.~M.~E.
  1975, \apj, 201, 624

\bibitem[{Landstreet} {et~al.}(1979)]{1979MNRAS.188..609L}
{Landstreet}, J.~D., {Borra}, E.~F., \& {Fontaine}, G. 1979, \mnras, 188, 609

\bibitem[{Landstreet} {et~al.}(2017)]{2017A&A...601A.129L}
{Landstreet}, J.~D., {Kochukhov}, O., {Alecian}, E., {et~al.} 2017, \aap, 601,
  A129

\bibitem[{Landstreet} {et~al.}(2008)]{2008A&A...481..465L}
{Landstreet}, J.~D., {Silaj}, J., {Andretta}, V., {et~al.} 2008, \aap, 481, 465

\bibitem[{Leone}(2007)]{2007MNRAS.382.1690L}
{Leone}, F. 2007, \mnras, 382, 1690

\bibitem[{Leone} \& {Catanzaro}(2001)]{2001A&A...365..118L}
{Leone}, F., \& {Catanzaro}, G. 2001, \aap, 365, 118

\bibitem[{Leone} \& {Kurtz}(2003)]{2003A&A...407L..67L}
{Leone}, F., \& {Kurtz}, D.~W. 2003, \aap, 407, L67

\bibitem[{Leone} {et~al.}(2003)]{2003A&A...405..223L}
{Leone}, F., {Plachinda}, S.~I., {Umana}, G., {Trigilio}, C., \& {Skulsky}, M.
  2003, \aap, 405, 223

\bibitem[{Martin} {et~al.}(2018)]{2018MNRAS.475.1521M}
{Martin}, A.~J., {Neiner}, C., {Oksala}, M.~E., {et~al.} 2018, \mnras, 475,
  1521

\bibitem[{Mathys}(1991)]{1991A&AS...89..121M}
{Mathys}, G. 1991, \aaps, 89, 121

\bibitem[{Mathys}(1994)]{1994A&AS..108..547M}
{Mathys}, G. 1994, \aaps, 108, 547

\bibitem[{Mathys}(2017)]{2017A&A...601A..14M}
{Mathys}, G. 2017, \aap, 601, A14

\bibitem[{Mathys} \& {Hubrig}(1995)]{1995A&A...293..810M}
{Mathys}, G., \& {Hubrig}, S. 1995, \aap, 293, 810

\bibitem[{Mathys} \& {Hubrig}(1997)]{1997A&AS..124..475M}
{Mathys}, G., \& {Hubrig}, S. 1997, \aaps, 124, 475

\bibitem[{McSwain}(2008)]{2008ApJ...686.1269M}
{McSwain}, M.~V. 2008, \apj, 686, 1269

\bibitem[{Monin} {et~al.}(2002)]{2002A&A...396..131M}
{Monin}, D.~N., {Fabrika}, S.~N., \& {Valyavin}, G.~G. 2002, \aap, 396, 131

\bibitem[{Morel} {et~al.}(2014)]{2014Msngr.157...27M}
{Morel}, T., {Castro}, N., {Fossati}, L., {et~al.} 2014, The Messenger, 157, 27

\bibitem[{Naz{\'e}} {et~al.}(2012)]{2012MNRAS.423.3413N}
{Naz{\'e}}, Y., {Bagnulo}, S., {Petit}, V., {et~al.} 2012, \mnras, 423, 3413

\bibitem[{Naz{\'e}} {et~al.}(2017)]{2017MNRAS.467..501N}
{Naz{\'e}}, Y., {Neiner}, C., {Grunhut}, J., {et~al.} 2017, \mnras, 467, 501

\bibitem[{Neiner} {et~al.}(2003{\natexlab{a}})]{2003A&A...406.1019N}
{Neiner}, C., {Geers}, V.~C., {Henrichs}, H.~F., {et~al.} 2003{\natexlab{a}},
  \aap, 406, 1019

\bibitem[{Neiner} {et~al.}(2003{\natexlab{b}})]{2003A&A...409..275N}
{Neiner}, C., {Hubert}, A.~M., {Fr{\'e}mat}, Y., {et~al.} 2003{\natexlab{b}},
  \aap, 409, 275

\bibitem[{Neiner} {et~al.}(2003{\natexlab{c}})]{2003A&A...411..565N}
{Neiner}, C., {Henrichs}, H.~F., {Floquet}, M., {et~al.} 2003{\natexlab{c}},
  \aap, 411, 565

\bibitem[{O'Toole} {et~al.}(2005)]{2005A&A...437..227O}
{O'Toole}, S.~J., {Jordan}, S., {Friedrich}, S., \& {Heber}, U. 2005, \aap,
  437, 227

\bibitem[{Parker}(1995)]{1995ApJ...440..415P}
{Parker}, E.~N. 1995, \apj, 440, 415

\bibitem[{Petit} {et~al.}(2005)]{2005MNRAS.361..837P}
{Petit}, P., {Donati}, J.~F., {Auri{\`e}re}, M., {et~al.} 2005, \mnras, 361,
  837

\bibitem[{Phan-Bao} {et~al.}(2006)]{2006ApJ...646L..73P}
{Phan-Bao}, N., {Mart{\'\i}n}, E.~L., {Donati}, J.-F., \& {Lim}, J. 2006,
  \apjl, 646, L73

\bibitem[{Pilachowski} {et~al.}(1974)]{1974A&A....37..275P}
{Pilachowski}, C.~A., {Bonsack}, W.~K., \& {Wolff}, S.~C. 1974, \aap, 37, 275

\bibitem[{Pillitteri} {et~al.}(2018)]{2018A&A...610L...3P}
{Pillitteri}, I., {Fossati}, L., {Castro Rodriguez}, N., {Oskinova}, L., \&
  {Wolk}, S.~J. 2018, \aap, 610, L3

\bibitem[{Plachinda}(2000)]{2000A&A...360..642P}
{Plachinda}, S.~I. 2000, \aap, 360, 642

\bibitem[{Plachinda} \& {Tarasova}(1999)]{1999ApJ...514..402P}
{Plachinda}, S.~I., \& {Tarasova}, T.~N. 1999, \apj, 514, 402

\bibitem[{Plachinda} \& {Tarasova}(2000)]{2000ApJ...533.1016P}
{Plachinda}, S.~I., \& {Tarasova}, T.~N. 2000, \apj, 533, 1016

\bibitem[{Preston}(1967)]{1967ApJ...150..871P}
{Preston}, G.~W. 1967, \apj, 150, 871

\bibitem[{Preston}(1969{\natexlab{a}})]{1969ApJ...156..967P}
{Preston}, G.~W. 1969{\natexlab{a}}, \apj, 156, 967

\bibitem[{Preston}(1969{\natexlab{b}})]{1969ApJ...158.1081P}
{Preston}, G.~W. 1969{\natexlab{b}}, \apj, 158, 1081

\bibitem[{Preston}(1969{\natexlab{c}})]{1969ApJ...158..243P}
{Preston}, G.~W. 1969{\natexlab{c}}, \apj, 158, 243

\bibitem[{Preston}(1969{\natexlab{d}})]{1969ApJ...158..251P}
{Preston}, G.~W. 1969{\natexlab{d}}, \apj, 158, 251

\bibitem[{Preston} \& {Stepien}(1968{\natexlab{a}})]{1968ApJ...151..583P}
{Preston}, G.~W., \& {Stepien}, K. 1968{\natexlab{a}}, \apj, 151, 583

\bibitem[{Preston} \& {Stepien}(1968{\natexlab{b}})]{1968ApJ...154..971P}
{Preston}, G.~W., \& {Stepien}, K. 1968{\natexlab{b}}, \apj, 154, 971

\bibitem[{Preston} \& {Stepien}(1968{\natexlab{c}})]{1968ApJ...151..577P}
{Preston}, G.~W., \& {Stepien}, K. 1968{\natexlab{c}}, \apj, 151, 577

\bibitem[{Preston} {et~al.}(1969)]{1969ApJ...156..653P}
{Preston}, G.~W., {Stepien}, K., \& {Wolff}, S.~C. 1969, \apj, 156, 653

\bibitem[{Rakos} {et~al.}(1977)]{1977A&A....56..453R}
{Rakos}, K.~D., {Schermann}, A., {Weiss}, W.~W., \& {Wood}, H.~J. 1977, \aap,
  56, 453

\bibitem[{Renson} \& {Manfroid}(2009)]{2009A&A...498..961R}
{Renson}, P., \& {Manfroid}, J. 2009, \aap, 498, 961

\bibitem[{Rivinius} {et~al.}(2010)]{2010MNRAS.405L..46R}
{Rivinius}, T., {Szeifert}, T., {Barrera}, L., {et~al.} 2010, \mnras, 405, L46

\bibitem[{Romanyuk}(2011)]{2011AN....332..965R}
{Romanyuk}, I.~I. 2011, Astronomische Nachrichten, 332, 965

\bibitem[{Romanyuk} \& {Kudryavtsev}(1998)]{1998CoSka..27..485R}
{Romanyuk}, I.~I., \& {Kudryavtsev}, D.~O. 1998, Contributions of the
  Astronomical Observatory Skalnate Pleso, 27, 485

\bibitem[{Romanyuk} {et~al.}(2017)]{2017AstBu..72..391R}
{Romanyuk}, I.~I., {Semenko}, E.~A., {Kudryavtsev}, D.~O., {Moiseeva}, A.~V.,
  \& {Yakunin}, I.~A. 2017, Astrophysical Bulletin, 72, 391

\bibitem[{Romanyuk} {et~al.}(2018)]{2018CoSka..48..208R}
{Romanyuk}, I.~I., {Semenko}, E.~A., {Kudryavtsev}, D.~O., \& {Yakunin}, I.~A.
  2018, Contributions of the Astronomical Observatory Skalnate Pleso, 48, 208

\bibitem[{Rudy} \& {Kemp}(1978)]{1978MNRAS.183..595R}
{Rudy}, R.~J., \& {Kemp}, J.~C. 1978, \mnras, 183, 595

\bibitem[{Ruediger} \& {Scholz}(1988)]{1988AN....309..181R}
{Ruediger}, G., \& {Scholz}, G. 1988, Astronomische Nachrichten, 309, 181

\bibitem[{Rustamov} \& {Khotnyanskij}(1980)]{1980SvAL....6..202R}
{Rustamov}, Y.~S., \& {Khotnyanskij}, A.~N. 1980, Soviet Astronomy Letters, 6,
  202

\bibitem[{Ryabchikova} {et~al.}(2006)]{2006A&A...445L..47R}
{Ryabchikova}, T., {Kochukhov}, O., {Kudryavtsev}, D., {et~al.} 2006, \aap,
  445, L47

\bibitem[{Ryabchikova} {et~al.}(2007)]{2007A&A...462.1103R}
{Ryabchikova}, T., {Sachkov}, M., {Weiss}, W.~W., {et~al.} 2007, \aap, 462,
  1103

\bibitem[{Sargent} {et~al.}(1967)]{1967ApJ...147.1185S}
{Sargent}, W. L.~W., {Sargent}, A.~I., \& {Strittmatter}, P.~A. 1967, \apj,
  147, 1185

\bibitem[{Schneider} {et~al.}(2019)]{2019Natur.574..211S}
{Schneider}, F. R.~N., {Ohlmann}, S.~T., {Podsiadlowski}, P., {et~al.} 2019,
  \nat, 574, 211

\bibitem[{Schnerr} {et~al.}(2006)]{2006A&A...452..969S}
{Schnerr}, R.~S., {Verdugo}, E., {Henrichs}, H.~F., \& {Neiner}, C. 2006, \aap,
  452, 969

\bibitem[{Sch{\"o}ller} {et~al.}(2017)]{2017A&A...599A..66S}
{Sch{\"o}ller}, M., {Hubrig}, S., {Fossati}, L., {et~al.} 2017, \aap, 599, A66

\bibitem[{Scholz}(1971)]{1971AN....292..279S}
{Scholz}, G. 1971, Astronomische Nachrichten, 292, 279

\bibitem[{Scholz}(1978)]{1978AN....299...81S}
{Scholz}, G. 1978, Astronomische Nachrichten, 299, 81

\bibitem[{Scholz}(1979)]{1979AN....300..213S}
{Scholz}, G. 1979, Astronomische Nachrichten, 300, 213

\bibitem[{Scholz}(1983)]{1983Ap&SS..94..159S}
{Scholz}, G. 1983, \apss, 94, 159

\bibitem[{Scholz} \& {Gerth}(1980)]{1980AN....301..211S}
{Scholz}, G., \& {Gerth}, E. 1980, Astronomische Nachrichten, 301, 211

\bibitem[{Schr{\"o}der} {et~al.}(2008)]{2008A&A...484..479S}
{Schr{\"o}der}, C., {Hubrig}, S., \& {Schmitt}, J.~H.~M.~M. 2008, \aap, 484,
  479

\bibitem[{Severny}(1970)]{1970ApJ...159L..73S}
{Severny}, A. 1970, \apjl, 159, L73

\bibitem[{Shen} {et~al.}(2023)]{2023RAA....23a5002S}
{Shen}, D.-X., {Liu}, J.-Z., {Zhu}, C.-H., {et~al.} 2023, Research in Astronomy
  and Astrophysics, 23, 015002

\bibitem[{Shore} {et~al.}(1990)]{1990ApJ...348..242S}
{Shore}, S.~N., {Brown}, D.~N., {Sonneborn}, G., {Landstreet}, J.~D., \&
  {Bohlender}, D.~A. 1990, \apj, 348, 242

\bibitem[{Shorlin} {et~al.}(2002)]{2002A&A...392..637S}
{Shorlin}, S.~L.~S., {Wade}, G.~A., {Donati}, J.~F., {et~al.} 2002, \aap, 392,
  637

\bibitem[{Shultz} {et~al.}(2019{\natexlab{a}})]{2019MNRAS.490.4154S}
{Shultz}, M.~E., {Johnston}, C., {Labadie-Bartz}, J., {et~al.}
  2019{\natexlab{a}}, \mnras, 490, 4154

\bibitem[{Shultz} {et~al.}(2018)]{2018pas8.conf..146S}
{Shultz}, M., {Wade}, G.~A., {Neiner}, C., \& {Kochukhov}, O. 2018, in 3rd
  BRITE Science Conference, ed. G.~A. {Wade}, D.~{Baade}, J.~A. {Guzik}, \&
  R.~{Smolec}, Vol.~8, 146

\bibitem[{Shultz} {et~al.}(2012)]{2012ApJ...750....2S}
{Shultz}, M., {Wade}, G.~A., {Grunhut}, J., {et~al.} 2012, \apj, 750, 2

\bibitem[{Shultz} {et~al.}(2019{\natexlab{b}})]{2019MNRAS.482.3950S}
{Shultz}, M., {Le Bouquin}, J.~B., {Rivinius}, T., {et~al.} 2019{\natexlab{b}},
  \mnras, 482, 3950

\bibitem[{Slovak}(1982)]{1982ApJ...262..282S}
{Slovak}, M.~H. 1982, \apj, 262, 282

\bibitem[{Smirnov} {et~al.}(2003)]{2003AstL...29..258S}
{Smirnov}, D.~A., {Lamzin}, S.~A., \& {Fabrika}, S.~N. 2003, Astronomy Letters,
  29, 258

\bibitem[{Tarasov} {et~al.}(1992)]{1992IBVS.3703....1T}
{Tarasov}, A.~E., {Bychkov}, V.~D., \& {Shtol}, V.~G. 1992, Information
  Bulletin on Variable Stars, 3703, 1

\bibitem[{Tarasova}(2002)]{2002ARep...46..474T}
{Tarasova}, T.~N. 2002, Astronomy Reports, 46, 474

\bibitem[{Thompson}(1983)]{1983MNRAS.205p..43T}
{Thompson}, I.~B. 1983, \mnras, 205, 43P

\bibitem[{Thompson} {et~al.}(1987)]{1987ApJS...64..219T}
{Thompson}, I.~B., {Brown}, D.~N., \& {Landstreet}, J.~D. 1987, \apjs, 64, 219

\bibitem[{Thompson} \& {Landstreet}(1985)]{1985ApJ...289L...9T}
{Thompson}, I.~B., \& {Landstreet}, J.~D. 1985, \apjl, 289, L9

\bibitem[{Valyavin} {et~al.}(2006)]{2006ApJ...648..559V}
{Valyavin}, G., {Bagnulo}, S., {Fabrika}, S., {et~al.} 2006, \apj, 648, 559

\bibitem[{Valyavin} {et~al.}(2003)]{2003ARep...47..587V}
{Valyavin}, G.~G., {Burlakova}, T.~E., {Fabrika}, S.~N., \& {Monin}, D.~N.
  2003, Astronomy Reports, 47, 587

\bibitem[{van den Heuvel}(1971)]{1971A&A....11..461V}
{van den Heuvel}, E.~P.~J. 1971, \aap, 11, 461

\bibitem[{Vetesnik}(1983)]{1983IBVS.2395....1V}
{Vetesnik}, M. 1983, Information Bulletin on Variable Stars, 2395, 1

\bibitem[{Vogt}(1980)]{1980ApJ...240..567V}
{Vogt}, S.~S. 1980, \apj, 240, 567

\bibitem[{Vogt} {et~al.}(1980)]{1980ApJ...236..308V}
{Vogt}, S.~S., {Tull}, R.~G., \& {Kelton}, P.~W. 1980, \apj, 236, 308

\bibitem[{Wade}(2004)]{2004IAUS..224..235W}
{Wade}, G.~A. 2004, in The A-Star Puzzle, ed. J.~{Zverko}, J.~{Ziznovsky},
  S.~J. {Adelman}, \& W.~W. {Weiss}, Vol. 224, 235

\bibitem[{Wade} {et~al.}(2007)]{2007MNRAS.376.1145W}
{Wade}, G.~A., {Bagnulo}, S., {Drouin}, D., {Landstreet}, J.~D., \& {Monin}, D.
  2007, \mnras, 376, 1145

\bibitem[{Wade} {et~al.}(1997{\natexlab{a}})]{1997A&A...320..172W}
{Wade}, G.~A., {Bohlender}, D.~A., {Brown}, D.~N., {et~al.} 1997{\natexlab{a}},
  \aap, 320, 172

\bibitem[{Wade} {et~al.}(2002)]{2002A&A...392L..17W}
{Wade}, G.~A., {Chadid}, M., {Shorlin}, S.~L.~S., {Bagnulo}, S., \& {Weiss},
  W.~W. 2002, \aap, 392, L17

\bibitem[{Wade} {et~al.}(2000{\natexlab{a}})]{2000MNRAS.313..851W}
{Wade}, G.~A., {Donati}, J.~F., {Landstreet}, J.~D., \& {Shorlin}, S.~L.~S.
  2000{\natexlab{a}}, \mnras, 313, 851

\bibitem[{Wade} {et~al.}(2006{\natexlab{a}})]{2006A&A...451..195W}
{Wade}, G.~A., {Fullerton}, A.~W., {Donati}, J.~F., {et~al.}
  2006{\natexlab{a}}, \aap, 451, 195

\bibitem[{Wade} {et~al.}(2012)]{2012ASPC..464..405W}
{Wade}, G.~A., {Grunhut}, J.~H., \& {MiMeS Collaboration}. 2012, in
  Astronomical Society of the Pacific Conference Series, Vol. 464,
  Circumstellar Dynamics at High Resolution, ed. A.~C. {Carciofi} \&
  T.~{Rivinius}, 405

\bibitem[{Wade} {et~al.}(2000{\natexlab{b}})]{2000A&A...355.1080W}
{Wade}, G.~A., {Kudryavtsev}, D., {Romanyuk}, I.~I., {Landstreet}, J.~D., \&
  {Mathys}, G. 2000{\natexlab{b}}, \aap, 355, 1080

\bibitem[{Wade} {et~al.}(1997{\natexlab{b}})]{1997MNRAS.292..748W}
{Wade}, G.~A., {Landstreet}, J.~D., {Elkin}, V.~G., \& {Romanyuk}, I.~I.
  1997{\natexlab{b}}, \mnras, 292, 748

\bibitem[{Wade} {et~al.}(1996{\natexlab{a}})]{1996A&A...307..500W}
{Wade}, G.~A., {Neagu}, E., \& {Landstreet}, J.~D. 1996{\natexlab{a}}, \aap,
  307, 500

\bibitem[{Wade} {et~al.}(1996{\natexlab{b}})]{1996A&A...314..491W}
{Wade}, G.~A., {North}, P., {Mathys}, G., \& {Hubrig}, S. 1996{\natexlab{b}},
  \aap, 314, 491

\bibitem[{Wade} {et~al.}(2005)]{2005A&A...442L..31W}
{Wade}, G.~A., {Drouin}, D., {Bagnulo}, S., {et~al.} 2005, \aap, 442, L31

\bibitem[{Wade} {et~al.}(2006{\natexlab{b}})]{2006A&A...451..293W}
{Wade}, G.~A., {Auri{\`e}re}, M., {Bagnulo}, S., {et~al.} 2006{\natexlab{b}},
  \aap, 451, 293

\bibitem[{Wade} {et~al.}(2006{\natexlab{c}})]{2006A&A...458..569W}
{Wade}, G.~A., {Smith}, M.~A., {Bohlender}, D.~A., {et~al.} 2006{\natexlab{c}},
  \aap, 458, 569

\bibitem[{Wade} {et~al.}(2016)]{2016MNRAS.456....2W}
{Wade}, G.~A., {Neiner}, C., {Alecian}, E., {et~al.} 2016, \mnras, 456, 2

\bibitem[{Wolff} \& {Wolff}(1976)]{1976ApJ...203..171W}
{Wolff}, R.~J., \& {Wolff}, S.~C. 1976, \apj, 203, 171

\bibitem[{Wolff}(1969{\natexlab{a}})]{1969ApJ...157..253W}
{Wolff}, S.~C. 1969{\natexlab{a}}, \apj, 157, 253

\bibitem[{Wolff}(1969{\natexlab{b}})]{1969ApJ...158.1231W}
{Wolff}, S.~C. 1969{\natexlab{b}}, \apj, 158, 1231

\bibitem[{Wolff}(1973)]{1973ApJ...186..951W}
{Wolff}, S.~C. 1973, \apj, 186, 951

\bibitem[{Wolff}(1975)]{1975ApJ...202..127W}
{Wolff}, S.~C. 1975, \apj, 202, 127

\bibitem[{Wolff}(1978)]{1978PASP...90..412W}
{Wolff}, S.~C. 1978, \pasp, 90, 412

\bibitem[{Wolff} \& {Bonsack}(1972)]{1972ApJ...176..425W}
{Wolff}, S.~C., \& {Bonsack}, W.~K. 1972, \apj, 176, 425

\bibitem[{Wolff} \& {Hagen}(1976)]{1976PASP...88..119W}
{Wolff}, S.~C., \& {Hagen}, W. 1976, \pasp, 88, 119

\bibitem[{Wolff} \& {Morrison}(1974)]{1974PASP...86..935W}
{Wolff}, S.~C., \& {Morrison}, N.~D. 1974, \pasp, 86, 935

\bibitem[{Wolff} \& {Preston}(1978)]{1978PASP...90..406W}
{Wolff}, S.~C., \& {Preston}, G.~W. 1978, \pasp, 90, 406

\bibitem[{Wolff} \& {Wolff}(1970)]{1970ApJ...160.1049W}
{Wolff}, S.~C., \& {Wolff}, R.~J. 1970, \apj, 160, 1049

\bibitem[{Wolff} \& {Wolff}(1972)]{1972ApJ...176..433W}
{Wolff}, S.~C., \& {Wolff}, R.~J. 1972, \apj, 176, 433

\bibitem[{Wolstencroft} {et~al.}(1981)]{1981MNRAS.195p..39W}
{Wolstencroft}, R.~D., {Smith}, R.~J., \& {Clarke}, D. 1981, \mnras, 195, 39P

\bibitem[{Wood} \& {Campusano}(1975)]{1975A&A....45..303W}
{Wood}, H.~J., \& {Campusano}, L.~B. 1975, \aap, 45, 303

\bibitem[{Yakunin} {et~al.}(2011)]{2011AN....332..974Y}
{Yakunin}, I., {Romanyuk}, I., {Kudryavtsev}, D., \& {Semenko}, E. 2011,
  Astronomische Nachrichten, 332, 974

\bibitem[{Ziznovsky} \& {Romanyuk}(1990)]{1990BAICz..41..118Z}
{Ziznovsky}, J., \& {Romanyuk}, I.~I. 1990, Bulletin of the Astronomical
  Institutes of Czechoslovakia, 41, 118

\bibitem[{Zverko} {et~al.}(1989)]{1989CoSka..18...71Z}
{Zverko}, J., {Bychkov}, V.~D., {Zi{\v{z}}{\v{n}}ovsk{\'y}}, J., \& {Hric}, L.
  1989, Contributions of the Astronomical Observatory Skalnate Pleso, 18, 71

\end{thebibliography}

\appendix

\section{Catalogue of magnetic stars}

\begin{table}
	\centering
	\bc
	\begin{minipage}[]{\textwidth}
		\caption{ The whole catalogue for 1784 magnetic stars.\label{catalogue}}
	\end{minipage}
	\setlength{\tabcolsep}{0.1pt}
	\tiny
	\renewcommand{\arraystretch}{0.5}
\begin{tabular}{>{\centering\arraybackslash}m{2.5cm}>{\centering\arraybackslash}p{1cm}>{\centering\arraybackslash}p{1cm}>{\centering\arraybackslash}p{1cm}>{\centering\arraybackslash}p{1.5cm}>{\centering\arraybackslash}p{1cm}>{\centering\arraybackslash}p{0.5cm}>{\centering\arraybackslash}p{1cm}>{\centering\arraybackslash}p{1cm}>{\centering\arraybackslash}p{1cm}>{\centering\arraybackslash}m{3.5cm}}
	\hline\noalign{\smallskip}
	Name& HD Number & RA & DEC & Sp type & $\left \langle B_{\rm e}\right \rangle$ & $\left \langle\sigma_{\rm e}\right \rangle$ & $\chi^2/n$ & Method & Observation & References\\
	&&(J2000)&(J2000)&&($G$)&($G$)&&&count&\\
	\hline\noalign{\smallskip}
HD 108 & 108 & 1.5141  & 63.6797  & O4-8f?p & 88  & 27  & 18.145  & all & 37 & \citeyear{2017MNRAS.465.2432G}\\
HD 166 & 166 & 1.6533  & 29.0215  & G0V & 2125  &  &  & metal & 1 & {\citeyear{1983A&AS...53..119D}}\\
* 4 Cet & 315 & 1.9338  & -2.5487  & ApSi & 1438  & 692  & 4.201  & metal & 4 & \citeyear{2006MNRAS.372.1804K}\\
* alf And & 358 & 2.0969  & 29.0904  & B8IV-VHgMn & 103  & 93  & 2.751  & all & 66 & \citeyear{2006AN....327..289H},\citeyear{1970ApJ...159L..73S},{\citeyear{2012A&A...538A.129B}},\citeyear{1973ApJ...185L.139B},\citeyear{1980ApJS...42..421B},{\citeyear{2015A&A...583A.115B}},{\citeyear{2006A&A...451..293W}}\\
HD 653 & 653 & 2.7162  & 40.5776  &  & 180  & 112  & 4.495  & all & 4 & \citeyear{2017AstBu..72..391R}\\
* gam Peg & 886 & 3.3090  & 15.1836  & B2IV & 694  & 369  & 2.183  & H1 Balmer & 12 & \citeyear{1982ApJ...258..639L},\citeyear{1978MNRAS.183..595R}\\
\hline

\end{tabular}
	    \begin{minipage}{\textwidth}
	Table continued on next page
		\end{minipage}

\ec
\end{table}

\begin{table}
	\centering
	\bc
	\begin{minipage}{\textwidth}
		\tiny{continued}
	\end{minipage}
	\setlength{\tabcolsep}{0.1pt}
	\tiny
	\renewcommand{\arraystretch}{0.5}
	\begin{tabular}{>{\centering\arraybackslash}m{2.5cm}>{\centering\arraybackslash}p{1cm}>{\centering\arraybackslash}p{1cm}>{\centering\arraybackslash}p{1cm}>{\centering\arraybackslash}p{1.5cm}>{\centering\arraybackslash}p{1cm}>{\centering\arraybackslash}p{0.5cm}>{\centering\arraybackslash}p{1cm}>{\centering\arraybackslash}p{1cm}>{\centering\arraybackslash}p{1cm}>{\centering\arraybackslash}m{3.5cm}}
		\hline\noalign{\smallskip}
	Name& HD Number & RA & DEC & Sp type & $\left \langle B_{\rm e}\right \rangle$ & $\left \langle\sigma_{\rm e}\right \rangle$ & $\chi^2/n$ & Method & Observation & References\\
		&&(J2000)&(J2000)&&($G$)&($G$)&&&count&\\
		\hline\noalign{\smallskip}

* del Eri & 23249 & 55.8121  & -9.7634  & K0+IV & 2  & 5  & 0.275  & metal & 3 & \citeyear{1984ApJ...284..211B},{\citeyear{1994A&A...291..890H}}\\
HD 23371 & 23371 & 58.1139  & 76.6092  & K0 & 13  & 7  & 5.638  & metal & 3 & {\citeyear{1994A&A...291..890H}}\\
* 20 Tau & 23408 & 56.4567  & 24.3677  & B8III & 98  & 102  & 1.340  & all & 5 & \citeyear{2006AN....327..289H},{\citeyear{2015A&A...583A.115B}},{\citeyear{2006A&A...450..763K}}\\
HD 23478 & 23478 & 56.6703  & 32.2902  & B3IV & 1322  & 90  & 263.045  & all & 4 & {\citeyear{2015A&A...578L...3H}}\\
HD 23598 & 23598 & 56.8370  & 25.5258  & F8 & 235  & 161  & 0.834  & all & 3 & {\citeyear{2015A&A...583A.115B}}\\
HD 23924 & 23924 & 57.4205  & 23.3416  & A5VkA7mA7 & 55  & 55  & 1.067  & all & 2 & \citeyear{2017AstBu..72..391R}\\
\hline
		
	\end{tabular}
	\begin{minipage}{\textwidth}
		Table continued on next page
	\end{minipage}
	
	\ec
\end{table}

\begin{table}
	\centering
	\bc
	\begin{minipage}{\textwidth}
		\tiny{continued}
	\end{minipage}
	\setlength{\tabcolsep}{0.1pt}
	\tiny
	\renewcommand{\arraystretch}{0.5}
	\begin{tabular}{>{\centering\arraybackslash}m{2.5cm}>{\centering\arraybackslash}p{1cm}>{\centering\arraybackslash}p{1cm}>{\centering\arraybackslash}p{1cm}>{\centering\arraybackslash}p{1.5cm}>{\centering\arraybackslash}p{1cm}>{\centering\arraybackslash}p{0.5cm}>{\centering\arraybackslash}p{1cm}>{\centering\arraybackslash}p{1cm}>{\centering\arraybackslash}p{1cm}>{\centering\arraybackslash}m{3.5cm}}
		\hline\noalign{\smallskip}
	Name& HD Number & RA & DEC & Sp type & $\left \langle B_{\rm e}\right \rangle$ & $\left \langle\sigma_{\rm e}\right \rangle$ & $\chi^2/n$ & Method & Observation & References\\
		&&(J2000)&(J2000)&&($G$)&($G$)&&&count&\\
		\hline\noalign{\smallskip}

V* V1093 Ori & 36313 & 82.6885  & -0.3734  & ApSi & 541  & 155  & 9.207  & all & 4 & \citeyear{2011AN....332..974Y},\citeyear{2017AstBu..72..391R},\citeyear{1981ApJ...249L..39B}\\
HD 36429 & 36429 & 82.9218  & 2.8328  & B7II & 425  & 170  & 6.250  & metal & 1 & \citeyear{1981ApJ...249L..39B}\\
* del02 Ori & 36485 & 83.0017  & -0.2845  & B2Vsn & 2776  & 254  & 164.570  & all & 19 & {\citeyear{1997A&AS..124..475M}},\citeyear{2017AstBu..72..391R},\citeyear{1987ApJ...323..325B}\\
* ups Ori & 36512 & 82.9828  & -7.3015  & O9.7V & 66  & 75  & 0.565  & all & 2 & \citeyear{2017MNRAS.465.2432G},\citeyear{1982ApJ...258..639L}\\
V* V1099 Ori & 36526 & 83.0545  & -1.6005  & B8II & 2447  & 258  & 106.661  & all & 4 & \citeyear{2011AN....332..974Y},\citeyear{2017AstBu..72..391R},\citeyear{1981ApJ...249L..39B}\\
HD 36540 & 36540 & 83.0618  & -4.5184  & B7IVp & 375  & 168  & 13.995  & all & 10 & \citeyear{2011AN....332..974Y},\citeyear{1981ApJ...249L..39B},{\citeyear{2006A&A...450..777B}},{\citeyear{2015A&A...583A.115B}},\citeyear{2017AstBu..72..391R}\\
\hline
		
	\end{tabular}
	\begin{minipage}{\textwidth}
		Table continued on next page
	\end{minipage}
	
	\ec
\end{table}

\begin{table}
	\centering
	\bc
	\begin{minipage}{\textwidth}
		\tiny{continued}
	\end{minipage}
	\setlength{\tabcolsep}{0.1pt}
	\tiny
	\renewcommand{\arraystretch}{0.5}
	\begin{tabular}{>{\centering\arraybackslash}m{2.5cm}>{\centering\arraybackslash}p{1cm}>{\centering\arraybackslash}p{1cm}>{\centering\arraybackslash}p{1cm}>{\centering\arraybackslash}p{1.5cm}>{\centering\arraybackslash}p{1cm}>{\centering\arraybackslash}p{0.5cm}>{\centering\arraybackslash}p{1cm}>{\centering\arraybackslash}p{1cm}>{\centering\arraybackslash}p{1cm}>{\centering\arraybackslash}m{3.5cm}}
		\hline\noalign{\smallskip}
	Name& HD Number & RA & DEC & Sp type & $\left \langle B_{\rm e}\right \rangle$ & $\left \langle\sigma_{\rm e}\right \rangle$ & $\chi^2/n$ & Method & Observation & References\\
		&&(J2000)&(J2000)&&($G$)&($G$)&&&count&\\
		\hline\noalign{\smallskip}

HD 47756 & 47756 & 100.1331  & 6.3713  & B8pSiCrSr & 404  & 117  & 13.228  & metal & 6 & \citeyear{2006MNRAS.372.1804K}\\
* 15 Mon & 47839 & 100.2444  & 9.8958  & O7V+B1.5/2V & 180  & 165  & 1.867  & all & 21 & {\citeyear{2013A&A...551A..33H}},{\citeyear{2015A&A...583A.115B}},\citeyear{2017MNRAS.465.2432G}\\
HD 48099 & 48099 & 100.4968  & 6.3454  & O5V((f))z+O9:V & 44  & 64  & 0.337  & all & 4 & \citeyear{2017MNRAS.465.2432G}\\
HD 48279 & 48279AA & 100.6689  & 1.7162  & O8V+F2V & 110  & 74  & 2.248  & all & 8 & {\citeyear{2015A&A...582A..45F}}\\
HD 48279 & 48279AB & 100.6689  & 1.7162  & O8V+F2V & 119  & 78  & 2.791  & all & 4 & {\citeyear{2015A&A...582A..45F}}\\
HD 48279 & 48279AC & 100.6689  & 1.7162  & O8V+F2V & 49  & 45  & 1.484  & all & 4 & {\citeyear{2015A&A...582A..45F}}\\
\hline
		
	\end{tabular}
	\begin{minipage}{\textwidth}
		Table continued on next page
	\end{minipage}
	
	\ec
\end{table}

\begin{table}
	\centering
	\bc
	\begin{minipage}{\textwidth}
		\tiny{continued}
	\end{minipage}
	\setlength{\tabcolsep}{0.1pt}
	\tiny
	\renewcommand{\arraystretch}{0.5}
	\begin{tabular}{>{\centering\arraybackslash}m{2.5cm}>{\centering\arraybackslash}p{1cm}>{\centering\arraybackslash}p{1cm}>{\centering\arraybackslash}p{1cm}>{\centering\arraybackslash}p{1.5cm}>{\centering\arraybackslash}p{1cm}>{\centering\arraybackslash}p{0.5cm}>{\centering\arraybackslash}p{1cm}>{\centering\arraybackslash}p{1cm}>{\centering\arraybackslash}p{1cm}>{\centering\arraybackslash}m{3.5cm}}
		\hline\noalign{\smallskip}
	Name& HD Number & RA & DEC & Sp type & $\left \langle B_{\rm e}\right \rangle$ & $\left \langle\sigma_{\rm e}\right \rangle$ & $\chi^2/n$ & Method & Observation & References\\
		&&(J2000)&(J2000)&&($G$)&($G$)&&&count&\\
		\hline\noalign{\smallskip}

HD 64486 & 64486 & 121.1961  & 79.4796  & B9.5IVs & 855  & 514  & 5.402  & all & 6 & {\citeyear{1993A&A...269..355B}},{\citeyear{1971A&A....11..461V}}\\
HD 64740 & 64740 & 118.2651  & -49.6130  & B2V & 572  & 114  & 24.992  & all & 21 & \citeyear{1979ApJ...228..809B},\citeyear{1987ApJ...323..325B}\\
* 53 Cam & 65339 & 120.4268  & 60.3244  & A3VpSrSiCrEu & 3669  & 445  & 592.692  & all & 101 & \citeyear{1978MNRAS.183..595R},\citeyear{2000MNRAS.313..851W},\citeyear{2017AstBu..72..391R},\citeyear{1977SvAL....3..273G},{\citeyear{2001A&A...365..118L}},\citeyear{1968ApJ...151..583P},\citeyear{1971AN....292..279S},\newline \citeyear{1973ApJ...184L..79A},\citeyear{1984ApJ...284..211B},{\citeyear{1996A&A...312L...5E}},\citeyear{1975ApJ...201..624L},\citeyear{1977ApJ...212..141B},\citeyear{1958ApJS....3..141B}\\
HD 65691 & 65691 & 119.0836  & -60.8010  & B8/9V & 267  & 175  & 2.782  & all & 6 & {\citeyear{2006A&A...450..777B}},{\citeyear{2015A&A...583A.115B}}\\
V* V607 Car & 65712 & 119.0945  & -61.2839  & ApSi & 940  & 73  & 178.015  & all & 12 & {\citeyear{2006A&A...450..777B}},{\citeyear{2015A&A...583A.115B}}\\
\hline
		
	\end{tabular}
	\begin{minipage}{\textwidth}
		Table continued on next page
	\end{minipage}
	
	\ec
\end{table}

\begin{table}
	\centering
	\bc
	\begin{minipage}{\textwidth}
		\tiny{continued}
	\end{minipage}
	\setlength{\tabcolsep}{0.1pt}
	\tiny
	\renewcommand{\arraystretch}{0.5}
	\begin{tabular}{>{\centering\arraybackslash}m{2.5cm}>{\centering\arraybackslash}p{1cm}>{\centering\arraybackslash}p{1cm}>{\centering\arraybackslash}p{1cm}>{\centering\arraybackslash}p{1.5cm}>{\centering\arraybackslash}p{1cm}>{\centering\arraybackslash}p{0.5cm}>{\centering\arraybackslash}p{1cm}>{\centering\arraybackslash}p{1cm}>{\centering\arraybackslash}p{1cm}>{\centering\arraybackslash}m{3.5cm}}
		\hline\noalign{\smallskip}
	Name& HD Number & RA & DEC & Sp type & $\left \langle B_{\rm e}\right \rangle$ & $\left \langle\sigma_{\rm e}\right \rangle$ & $\chi^2/n$ & Method & Observation & References\\
		&&(J2000)&(J2000)&&($G$)&($G$)&&&count&\\
		\hline\noalign{\smallskip}

* mu. UMa & 89758 & 155.5822  & 41.4995  & M0III & 4  & 11  & 0.137  & metal & 1 & \citeyear{1984ApJ...284..211B}\\
V* ET UMa & 89822 & 156.0327  & 65.5664  & A1:VpSiSrHg & 639  & 212  & 20.728  & all & 13 & \citeyear{1970ApJ...160.1077C},{\citeyear{1993A&A...269..355B}},\citeyear{1980ApJS...42..421B},\citeyear{1958ApJS....3..141B}\\
HD 89856 & 89856 & 155.2305  & -51.8616  & B9 & 211  & 109  & 4.138  & all & 6 & {\citeyear{2006A&A...450..777B}},{\citeyear{2015A&A...583A.115B}}\\
HD 89900 & 89900 & 155.3076  & -51.7146  & A0/1IV/V & 86  & 134  & 0.283  & all & 3 & {\citeyear{2015A&A...583A.115B}}\\
V* V343 Vel & 89901 & 155.2970  & -51.7490  & B8/9II/IV & 202  & 149  & 1.460  & all & 6 & {\citeyear{2006A&A...450..777B}},{\citeyear{2015A&A...583A.115B}}\\
HD 89915 & 89915 & 155.3328  & -51.7413  & B9/A0V & 168  & 128  & 1.927  & all & 6 & {\citeyear{2006A&A...450..777B}},{\citeyear{2015A&A...583A.115B}}\\
\hline
		
	\end{tabular}
	\begin{minipage}{\textwidth}
		Table continued on next page
	\end{minipage}
	
	\ec
\end{table}

\begin{table}
	\centering
	\bc
	\begin{minipage}{\textwidth}
		\tiny{continued}
	\end{minipage}
	\setlength{\tabcolsep}{0.1pt}
	\tiny
	\renewcommand{\arraystretch}{0.5}
	\begin{tabular}{>{\centering\arraybackslash}m{2.5cm}>{\centering\arraybackslash}p{1cm}>{\centering\arraybackslash}p{1cm}>{\centering\arraybackslash}p{1cm}>{\centering\arraybackslash}p{1.5cm}>{\centering\arraybackslash}p{1cm}>{\centering\arraybackslash}p{0.5cm}>{\centering\arraybackslash}p{1cm}>{\centering\arraybackslash}p{1cm}>{\centering\arraybackslash}p{1cm}>{\centering\arraybackslash}m{3.5cm}}
		\hline\noalign{\smallskip}
	Name& HD Number & RA & DEC & Sp type & $\left \langle B_{\rm e}\right \rangle$ & $\left \langle\sigma_{\rm e}\right \rangle$ & $\chi^2/n$ & Method & Observation & References\\
		&&(J2000)&(J2000)&&($G$)&($G$)&&&count&\\
		\hline\noalign{\smallskip}

HD 107696 & 107696 & 185.7060  & -57.6761  & B7Vn & 726  & 594  & 0.577  & all & 10 & \citeyear{2006AN....327..289H},{\citeyear{2015A&A...583A.115B}},{\citeyear{2006A&A...450..763K}}\\
NGC 4361 & 107969 & 186.1281  & -18.7849  & [WC] & 544  & 521  & 1.410  & all & 12 & {\citeyear{2015A&A...583A.115B}},{\citeyear{2012A&A...542A..64J}}\\
* 14 Com & 108283 & 186.6003  & 27.2682  & F1IV:npS\_sh & 163  & 190  & 0.733  & metal & 3 & \citeyear{2006MNRAS.372.1804K}\\
V* FT Vir & 108506 & 186.9648  & -4.6153  & F5IVn & 369  & 362  & 2.048  & all & 6 & \citeyear{2017AstBu..72..391R}\\
HD 108642 & 108642 & 187.1590  & 26.2270  & A2m & 13  & 18  & 0.576  & LSD & 2 & {\citeyear{2002A&A...392..637S}}\\
HD 108651 & 108651 & 187.1857  & 25.8993  & kA2hA9VmF0 & 346  & 220  & 1.814  & all & 6 & {\citeyear{2002A&A...392..637S}},{\citeyear{1971A&A....11..461V}}\\
\hline
		
	\end{tabular}
	\begin{minipage}{\textwidth}
		Table continued on next page
	\end{minipage}
	
	\ec
\end{table}

\begin{table}
	\centering
	\bc
	\begin{minipage}{\textwidth}
		\tiny{continued}
	\end{minipage}
	\setlength{\tabcolsep}{0.1pt}
	\tiny
	\renewcommand{\arraystretch}{0.5}
	\begin{tabular}{>{\centering\arraybackslash}m{2.5cm}>{\centering\arraybackslash}p{1cm}>{\centering\arraybackslash}p{1cm}>{\centering\arraybackslash}p{1cm}>{\centering\arraybackslash}p{1.5cm}>{\centering\arraybackslash}p{1cm}>{\centering\arraybackslash}p{0.5cm}>{\centering\arraybackslash}p{1cm}>{\centering\arraybackslash}p{1cm}>{\centering\arraybackslash}p{1cm}>{\centering\arraybackslash}m{3.5cm}}
		\hline\noalign{\smallskip}
	Name& HD Number & RA & DEC & Sp type & $\left \langle B_{\rm e}\right \rangle$ & $\left \langle\sigma_{\rm e}\right \rangle$ & $\chi^2/n$ & Method & Observation & References\\
		&&(J2000)&(J2000)&&($G$)&($G$)&&&count&\\
		\hline\noalign{\smallskip}

HD 128585 & 128585 & 219.8702  & -51.1580  & B3IV & 51  & 76  & 0.495  & all & 3 & {\citeyear{2015A&A...583A.115B}}\\
V* IT Lup & 128775 & 220.0803  & -45.7938  & ApSi & 332  & 96  & 21.606  & all & 6 & \citeyear{2006AN....327..289H},{\citeyear{2015A&A...583A.115B}},\citeyear{1987ApJS...64..219T},{\citeyear{2006A&A...450..763K}}\\
* alf Cir & 128898 & 220.6267  & -64.9751  & A7VpSrCrEu & 498  & 244  & 15.812  & all & 63 & \citeyear{1975PASP...87..961B},{\citeyear{2004A&A...415..661H}},{\citeyear{1994A&AS..108..547M}},{\citeyear{1991A&AS...89..121M}},{\citeyear{1997A&AS..124..475M}},{\citeyear{2015A&A...583A.115B}},{\citeyear{1975A&A....45..303W}},\newline \citeyear{1980ApJS...42..421B}\\
HD 128974 & 128974 & 220.2558  & -36.1349  & B9.5III & 176  & 136  & 1.126  & all & 9 & \citeyear{2006AN....327..289H},{\citeyear{2015A&A...583A.115B}},\citeyear{1987ApJS...64..219T},{\citeyear{2006A&A...450..763K}}\\
* pi.01 Boo & 129174 & 220.1815  & 16.4183  & B9IIIpHgMnSi & 102  & 62  & 2.387  & all & 4 & {\citeyear{2002A&A...392..637S}},\citeyear{1958ApJS....3..141B}\\
V* BU Cir & 129557 & 221.2957  & -55.6016  & B2III & 90  & 50  & 3.061  & all & 3 & {\citeyear{2015A&A...583A.115B}}\\
\hline
		
	\end{tabular}
	\begin{minipage}{\textwidth}
		Table continued on next page
	\end{minipage}
	
	\ec
\end{table}

\begin{table}
	\centering
	\bc
	\begin{minipage}{\textwidth}
		\tiny{continued}
	\end{minipage}
	\setlength{\tabcolsep}{0.1pt}
	\tiny
	\renewcommand{\arraystretch}{0.5}
	\begin{tabular}{>{\centering\arraybackslash}m{2.5cm}>{\centering\arraybackslash}p{1cm}>{\centering\arraybackslash}p{1cm}>{\centering\arraybackslash}p{1cm}>{\centering\arraybackslash}p{1.5cm}>{\centering\arraybackslash}p{1cm}>{\centering\arraybackslash}p{0.5cm}>{\centering\arraybackslash}p{1cm}>{\centering\arraybackslash}p{1cm}>{\centering\arraybackslash}p{1cm}>{\centering\arraybackslash}m{3.5cm}}
		\hline\noalign{\smallskip}
	Name& HD Number & RA & DEC & Sp type & $\left \langle B_{\rm e}\right \rangle$ & $\left \langle\sigma_{\rm e}\right \rangle$ & $\chi^2/n$ & Method & Observation & References\\
		&&(J2000)&(J2000)&&($G$)&($G$)&&&count&\\
		\hline\noalign{\smallskip}

HD 146001 & 146001 & 243.7226  & -25.4770  & B8V & 647  & 382  & 1.619  & H1 Balmer & 5 & \citeyear{1983ApJS...53..151B}\\
* del Oph & 146051 & 243.5864  & -3.6943  & M0.5III & 35  & 24  & 1.883  & all & 3 & \citeyear{1984ApJ...284..211B},\citeyear{2017MNRAS.465.2432G}\\
HD 146484 & 146484 & 244.9028  & -57.9349  & B6Ve & 222  & 165  & 3.453  & all & 6 & {\citeyear{2006A&A...450..777B}},{\citeyear{2015A&A...583A.115B}}\\
HD 146555 & 146555 & 244.9973  & -57.9078  & ApSi & 409  & 129  & 12.720  & all & 6 & {\citeyear{2006A&A...450..777B}},{\citeyear{2015A&A...583A.115B}}\\
V* V933 Sco & 147010 & 245.0229  & -20.0564  & B9II/III & 3979  & 393  & 141.112  & all & 56 & {\citeyear{1994A&AS..108..547M}},{\citeyear{1991A&AS...89..121M}},\citeyear{1982SvAL....8...12G},{\citeyear{1997A&AS..124..475M}},\citeyear{1987ApJS...64..219T}\\
* omi Sco & 147084 & 245.1591  & -24.1693  & A5II & 124  & 68  & 4.231  & all & 10 & {\citeyear{2015A&A...583A.115B}},{\citeyear{2012A&A...538A.129B}},\citeyear{1987ApJS...64..219T},{\citeyear{2008A&A...484..479S}}\\
\hline
		
	\end{tabular}
	\begin{minipage}{\textwidth}
		Table continued on next page
	\end{minipage}
	
	\ec
\end{table}

\begin{table}
	\centering
	\bc
	\begin{minipage}{\textwidth}
		\tiny{continued}
	\end{minipage}
	\setlength{\tabcolsep}{0.1pt}
	\tiny
	\renewcommand{\arraystretch}{0.5}
	\begin{tabular}{>{\centering\arraybackslash}m{2.5cm}>{\centering\arraybackslash}p{1cm}>{\centering\arraybackslash}p{1cm}>{\centering\arraybackslash}p{1cm}>{\centering\arraybackslash}p{1.5cm}>{\centering\arraybackslash}p{1cm}>{\centering\arraybackslash}p{0.5cm}>{\centering\arraybackslash}p{1cm}>{\centering\arraybackslash}p{1cm}>{\centering\arraybackslash}p{1cm}>{\centering\arraybackslash}m{3.5cm}}
		\hline\noalign{\smallskip}
	Name& HD Number & RA & DEC & Sp type & $\left \langle B_{\rm e}\right \rangle$ & $\left \langle\sigma_{\rm e}\right \rangle$ & $\chi^2/n$ & Method & Observation & References\\
		&&(J2000)&(J2000)&&($G$)&($G$)&&&count&\\
		\hline\noalign{\smallskip}

V* V539 Ara & 161783 & 267.6183  & -53.6124  & B2/3Vnn & 166  & 83  & 5.984  & all & 18 & {\citeyear{2015A&A...583A.115B}},\citeyear{2006MNRAS.369L..61H},{\citeyear{2012A&A...538A.129B}},\citeyear{2007ASPC..361..434H}\\
* mu.01 Her & 161797 & 266.6148  & 27.7207  & G5IV & 64  & 70  & 0.836  & metal & 1 & \citeyear{1983IAUS..102...35B}\\
HD 161817 & 161817 & 266.6693  & 25.7492  & sdA2 & 325  & 425  & 4.311  & metal & 4 & \citeyear{1998CoSka..27..452E}\\
* gam Oph & 161868 & 266.9732  & 2.7073  & A1VnkA0mA0 & 248  & 292  & 0.599  & H1 Balmer & 2 & {\citeyear{2002A&A...396..131M}}\\
HD 162305 & 162305 & 267.9703  & -34.7824  & B8/9III/IV & 143  & 105  & 2.301  & all & 9 & {\citeyear{2006A&A...450..777B}},{\citeyear{2015A&A...583A.115B}}\\
HD 162305 & 162305North & 267.9703  & -34.7824  & B8/9III/IV & 76  & 109  & 0.238  & all & 3 & {\citeyear{2015A&A...583A.115B}}\\
\hline
		
	\end{tabular}
	\begin{minipage}{\textwidth}
		Table continued on next page
	\end{minipage}
	
	\ec
\end{table}

\begin{table}
	\centering
	\bc
	\begin{minipage}{\textwidth}
		\tiny{continued}
	\end{minipage}
	\setlength{\tabcolsep}{0.1pt}
	\tiny
	\renewcommand{\arraystretch}{0.5}
	\begin{tabular}{>{\centering\arraybackslash}m{2.5cm}>{\centering\arraybackslash}p{1cm}>{\centering\arraybackslash}p{1cm}>{\centering\arraybackslash}p{1cm}>{\centering\arraybackslash}p{1.5cm}>{\centering\arraybackslash}p{1cm}>{\centering\arraybackslash}p{0.5cm}>{\centering\arraybackslash}p{1cm}>{\centering\arraybackslash}p{1cm}>{\centering\arraybackslash}p{1cm}>{\centering\arraybackslash}m{3.5cm}}
		\hline\noalign{\smallskip}
	Name& HD Number & RA & DEC & Sp type & $\left \langle B_{\rm e}\right \rangle$ & $\left \langle\sigma_{\rm e}\right \rangle$ & $\chi^2/n$ & Method & Observation & References\\
		&&(J2000)&(J2000)&&($G$)&($G$)&&&count&\\
		\hline\noalign{\smallskip}

* zet Aql & 177724 & 286.3525  & 13.8635  & A0IV-Vnn & 326  & 316  & 0.978  & H1 Balmer & 2 & {\citeyear{2002A&A...396..131M}}\\
HD 177863 & 177863 & 286.7847  & -18.7382  & B8V & 39  & 57  & 1.024  & all & 9 & {\citeyear{2015A&A...583A.115B}},\citeyear{2006MNRAS.369L..61H},\citeyear{2007ASPC..361..434H}\\
HD 178308 & 178308 & 286.7380  & 27.7619  &  & 154  & 176  & 0.661  & all & 4 & \citeyear{2017AstBu..72..391R}\\
HD 178892 & 178892 & 287.4781  & 14.9662  & ApSrCrEu & 5559  & 451  & 195.390  & all & 41 & {\citeyear{2006A&A...445L..47R}},\citeyear{2017AstBu..72..391R},\citeyear{2003AstL...29..400E},\citeyear{2006MNRAS.372.1804K}\\
HD 179218 & 179218 & 287.7969  & 15.7877  & A0Ve & 84  & 48  & 4.331  & all & 3 & {\citeyear{2015A&A...583A.115B}}\\
* 19 Lyr & 179527 & 287.9417  & 31.2835  & B8IIIpSiSr & 124  & 189  & 5.308  & all & 15 & {\citeyear{2007A&A...475.1053A}},\citeyear{1980ApJS...42..421B}\\
\hline
		
	\end{tabular}
	\begin{minipage}{\textwidth}
		Table continued on next page
	\end{minipage}
	
	\ec
\end{table}

\begin{table}
	\centering
	\bc
	\begin{minipage}{\textwidth}
		\tiny{continued}
	\end{minipage}
	\setlength{\tabcolsep}{0.1pt}
	\tiny
	\renewcommand{\arraystretch}{0.5}
	\begin{tabular}{>{\centering\arraybackslash}m{2.5cm}>{\centering\arraybackslash}p{1cm}>{\centering\arraybackslash}p{1cm}>{\centering\arraybackslash}p{1cm}>{\centering\arraybackslash}p{1.5cm}>{\centering\arraybackslash}p{1cm}>{\centering\arraybackslash}p{0.5cm}>{\centering\arraybackslash}p{1cm}>{\centering\arraybackslash}p{1cm}>{\centering\arraybackslash}p{1cm}>{\centering\arraybackslash}m{3.5cm}}
		\hline\noalign{\smallskip}
	Name& HD Number & RA & DEC & Sp type & $\left \langle B_{\rm e}\right \rangle$ & $\left \langle\sigma_{\rm e}\right \rangle$ & $\chi^2/n$ & Method & Observation & References\\
		&&(J2000)&(J2000)&&($G$)&($G$)&&&count&\\
		\hline\noalign{\smallskip}

HD 204827 & 204827 & 322.2407  & 58.7398  & O9.5IV & 76  & 52  & 2.347  & all & 2 & \citeyear{2017MNRAS.465.2432G}\\
* bet Cep & 205021 & 322.1650  & 70.5607  & B0.5IIIs & 1034  & 1540  & 2.856  & H1 Balmer & 10 & \citeyear{1978MNRAS.183..595R}\\
HD 205073 & 205073 & 322.8466  & 48.3565  & A1p & 6  & 18  & 0.108  & H1 Balmer & 2 & {\citeyear{2008A&A...481..465L}}\\
HD 205087 & 205087 & 323.1126  & 23.3946  & A0VpSiSr & 403  & 325  & 1.662  & all & 11 & {\citeyear{1993A&A...269..355B}},\citeyear{2006MNRAS.372.1804K}\\
HD 205331 & 205331 & 323.2921  & 48.3034  & A1III & 219  & 203  & 2.820  & all & 6 & {\citeyear{2008A&A...481..465L}},\citeyear{2006MNRAS.372.1804K}\\
HD 205805 & 205805 & 324.7942  & -46.0976  & sdB & 106  & 153  & 0.672  & all & 3 & {\citeyear{2015A&A...583A.115B}}\\
\hline
		
	\end{tabular}
	\begin{minipage}{\textwidth}
		Table continued on next page
	\end{minipage}
	
	\ec
\end{table}

\begin{table}
	\centering
	\bc
	\begin{minipage}{\textwidth}
		\tiny{continued}
	\end{minipage}
	\setlength{\tabcolsep}{0.1pt}
	\tiny
	\renewcommand{\arraystretch}{0.5}
	\begin{tabular}{>{\centering\arraybackslash}m{2.5cm}>{\centering\arraybackslash}p{1cm}>{\centering\arraybackslash}p{1cm}>{\centering\arraybackslash}p{1cm}>{\centering\arraybackslash}p{1.5cm}>{\centering\arraybackslash}p{1cm}>{\centering\arraybackslash}p{0.5cm}>{\centering\arraybackslash}p{1cm}>{\centering\arraybackslash}p{1cm}>{\centering\arraybackslash}p{1cm}>{\centering\arraybackslash}m{3.5cm}}
		\hline\noalign{\smallskip}
	Name& HD Number & RA & DEC & Sp type & $\left \langle B_{\rm e}\right \rangle$ & $\left \langle\sigma_{\rm e}\right \rangle$ & $\chi^2/n$ & Method & Observation & References\\
		&&(J2000)&(J2000)&&($G$)&($G$)&&&count&\\
		\hline\noalign{\smallskip}

V* V917 Ori & 252248 & 92.1140  & 13.9309  & B3Vn & 141  & 92  & 2.569  & all & 6 & {\citeyear{2006A&A...450..777B}},{\citeyear{2015A&A...583A.115B}}\\
HD 252266 & 252266 & 92.1257  & 13.9590  & B3V & 88  & 73  & 1.242  & all & 6 & {\citeyear{2006A&A...450..777B}},{\citeyear{2015A&A...583A.115B}}\\
HD 258686 & 258686 & 97.6961  & 10.0629  & A0IIIp\_sh? & 5966  & 319  & 370.699  & all & 7 & \citeyear{2017AstBu..72..391R},\citeyear{2006MNRAS.372.1804K}\\
HD 258691 & 258691 & 97.6388  & 4.6910  & O9V & 52  & 36  & 2.086  & all & 1 & \citeyear{2017MNRAS.465.2432G}\\
HD 259012 & 259012 & 97.8894  & 4.8444  & B1Vn+B4/A1Ve & 119  & 98  & 1.135  & all & 8 & {\citeyear{2015A&A...582A..45F}}\\
HD 259105 & 259105 & 97.9667  & 4.9326  & B1V & 157  & 88  & 2.568  & all & 8 & {\citeyear{2015A&A...582A..45F}}\\
\hline
		
	\end{tabular}
	\begin{minipage}{\textwidth}
		Table continued on next page
	\end{minipage}
	
	\ec
\end{table}

\begin{table}
	\centering
	\bc
	\begin{minipage}{\textwidth}
		\tiny{continued}
	\end{minipage}
	\setlength{\tabcolsep}{0.1pt}
	\tiny
	\renewcommand{\arraystretch}{0.5}
	\begin{tabular}{>{\centering\arraybackslash}m{2.5cm}>{\centering\arraybackslash}p{1cm}>{\centering\arraybackslash}p{1cm}>{\centering\arraybackslash}p{1cm}>{\centering\arraybackslash}p{1.5cm}>{\centering\arraybackslash}p{1cm}>{\centering\arraybackslash}p{0.5cm}>{\centering\arraybackslash}p{1cm}>{\centering\arraybackslash}p{1cm}>{\centering\arraybackslash}p{1cm}>{\centering\arraybackslash}m{3.5cm}}
		\hline\noalign{\smallskip}
	Name& HD Number & RA & DEC & Sp type & $\left \langle B_{\rm e}\right \rangle$ & $\left \langle\sigma_{\rm e}\right \rangle$ & $\chi^2/n$ & Method & Observation & References\\
		&&(J2000)&(J2000)&&($G$)&($G$)&&&count&\\
		\hline\noalign{\smallskip}

BD+40 175B &  & 12.7916  & 41.1979  & A2 & 1603  & 170  & 111.044  & metal & 5 & \citeyear{1999AstL...25..809E}\\
BD+46 570 &  & 36.7488  & 46.8217  &  & 478  & 103  & 22.412  & metal & 3 & \citeyear{2006MNRAS.372.1804K}\\
BD+49 3789 &  & 333.7975  & 49.7585  & B7III & 561  & 400  & 1.967  & H1 Balmer & 1 & {\citeyear{2008A&A...481..465L}}\\
BD+53 1183 &  & 116.7748  & 53.6387  & ApSrCr & 643  & 246  & 40.626  & all & 5 & \citeyear{2017AstBu..72..391R}\\
BD+60 499 &  & 38.0698  & 61.5542  & O9.5V & 5  & 45  & 0.012  & all & 1 & \citeyear{2017MNRAS.465.2432G}\\
BD+75 325 &  & 122.7062  & 74.9661  & sdO5 & 1428  & 352  & 473.360  & all & 7 & \citeyear{1998CoSka..27..452E},{\citeyear{1996A&A...312L...5E}}\\
\hline
		
	\end{tabular}
	\begin{minipage}{\textwidth}
		Table continued on next page
	\end{minipage}
	
	\ec
\end{table}

\begin{table}
	\centering
	\bc
	\begin{minipage}{\textwidth}
		\tiny{continued}
	\end{minipage}
	\setlength{\tabcolsep}{0.1pt}
	\tiny
	\renewcommand{\arraystretch}{0.5}
	\begin{tabular}{>{\centering\arraybackslash}m{2.5cm}>{\centering\arraybackslash}p{1cm}>{\centering\arraybackslash}p{1cm}>{\centering\arraybackslash}p{1cm}>{\centering\arraybackslash}p{1.5cm}>{\centering\arraybackslash}p{1cm}>{\centering\arraybackslash}p{0.5cm}>{\centering\arraybackslash}p{1cm}>{\centering\arraybackslash}p{1cm}>{\centering\arraybackslash}p{1cm}>{\centering\arraybackslash}m{3.5cm}}
		\hline\noalign{\smallskip}
	Name& HD Number & RA & DEC & Sp type & $\left \langle B_{\rm e}\right \rangle$ & $\left \langle\sigma_{\rm e}\right \rangle$ & $\chi^2/n$ & Method & Observation & References\\
		&&(J2000)&(J2000)&&($G$)&($G$)&&&count&\\
		\hline\noalign{\smallskip}

Cl* NGC 2516 SBL 658 &  & 119.5481  & -60.7633  &  & 499  & 513  & 0.810  & all & 3 & {\citeyear{2015A&A...583A.115B}}\\
Cl* NGC 3114 AR 109 &  & 150.6531  & -59.9518  &  & 2409  & 809  & 13.820  & all & 3 & {\citeyear{2015A&A...583A.115B}}\\
Cl* NGC 3228 SC 15 &  & 155.3454  & -51.8378  &  & 694  & 512  & 1.351  & all & 3 & {\citeyear{2015A&A...583A.115B}}\\
Cl* NGC 346 ELS 100 &  & 14.8363  & -72.0606  & B1.5V & 1210  & 928  & 1.072  & all & 2 & {\citeyear{2017A&A...601A.136B}},{\citeyear{2020A&A...635A.163B}}\\
Cl* NGC 346 ELS 103 &  & 14.8370  & -72.0496  & B0.5V & 230  & 855  & 0.224  & all & 2 & {\citeyear{2017A&A...601A.136B}},{\citeyear{2020A&A...635A.163B}}\\
Cl* NGC 346 ELS 19 &  & 14.7732  & -72.1340  & A0II & 615  & 315  & 3.812  & all & 1 & {\citeyear{2020A&A...635A.163B}}\\
\hline
		
	\end{tabular}
	\begin{minipage}{\textwidth}
		Table continued on next page
	\end{minipage}
	
	\ec
\end{table}

\begin{table}
	\centering
	\bc
	\begin{minipage}{\textwidth}
		\tiny{continued}
	\end{minipage}
	\setlength{\tabcolsep}{0.1pt}
	\tiny
	\renewcommand{\arraystretch}{0.5}
	\begin{tabular}{>{\centering\arraybackslash}m{2.5cm}>{\centering\arraybackslash}p{1cm}>{\centering\arraybackslash}p{1cm}>{\centering\arraybackslash}p{1cm}>{\centering\arraybackslash}p{1.5cm}>{\centering\arraybackslash}p{1cm}>{\centering\arraybackslash}p{0.5cm}>{\centering\arraybackslash}p{1cm}>{\centering\arraybackslash}p{1cm}>{\centering\arraybackslash}p{1cm}>{\centering\arraybackslash}m{3.5cm}}
		\hline\noalign{\smallskip}
	Name& HD Number & RA & DEC & Sp type & $\left \langle B_{\rm e}\right \rangle$ & $\left \langle\sigma_{\rm e}\right \rangle$ & $\chi^2/n$ & Method & Observation & References\\
		&&(J2000)&(J2000)&&($G$)&($G$)&&&count&\\
		\hline\noalign{\smallskip}

LS 181 &  & 107.0545  & -10.6007  & A & 205  & 180  & 1.823  & all & 4 & {\citeyear{2006A&A...450..777B}},{\citeyear{2015A&A...583A.115B}}\\
LS IV -12 1 &  & 245.9333  & -12.2093  & sdO & 417  & 297  & 1.678  & all & 3 & {\citeyear{2015A&A...583A.115B}}\\
MACHO 5.5377.4508 &  & 78.3616  & -69.3654  & B1:II & 503  & 259  & 293.348  & all & 2 & {\citeyear{2017A&A...601A.136B}}\\
Mrk 51 &  & 186.0605  & 4.2260  &  & 507  & 250  & 2.769  & all & 4 & {\citeyear{2020A&A...635A.163B}}\\
NAME LMC &  & 80.8942  & -69.7561  &  & 324  & 256  & 1.881  & all & 2 & {\citeyear{2020A&A...635A.163B}}\\
NAME SMC 159-2 &  & 12.4947  & -73.3246  & O8f?p & 1410  & 593  & 3.939  & all & 5 & {\citeyear{2017A&A...601A.136B}},{\citeyear{2020A&A...635A.163B}}\\
\hline
		
	\end{tabular}
	\begin{minipage}{\textwidth}
		Table continued on next page
	\end{minipage}
	
	\ec
\end{table}

\begin{table}
	\centering
	\bc
	\begin{minipage}{\textwidth}
		\tiny{continued}
	\end{minipage}
	\setlength{\tabcolsep}{0.1pt}
	\tiny
	\renewcommand{\arraystretch}{0.5}
	\begin{tabular}{>{\centering\arraybackslash}m{2.5cm}>{\centering\arraybackslash}p{1cm}>{\centering\arraybackslash}p{1cm}>{\centering\arraybackslash}p{1cm}>{\centering\arraybackslash}p{1.5cm}>{\centering\arraybackslash}p{1cm}>{\centering\arraybackslash}p{0.5cm}>{\centering\arraybackslash}p{1cm}>{\centering\arraybackslash}p{1cm}>{\centering\arraybackslash}p{1cm}>{\centering\arraybackslash}m{3.5cm}}
		\hline\noalign{\smallskip}
	Name& HD Number & RA & DEC & Sp type & $\left \langle B_{\rm e}\right \rangle$ & $\left \langle\sigma_{\rm e}\right \rangle$ & $\chi^2/n$ & Method & Observation & References\\
		&&(J2000)&(J2000)&&($G$)&($G$)&&&count&\\
		\hline\noalign{\smallskip}

V* RX For &  & 47.8051  & -26.4830  & kA9hF6 & 62  & 82  & 0.389  & all & 3 & {\citeyear{2015A&A...583A.115B}}\\
V* SS For &  & 31.9666  & -26.8660  & kF5hF8 & 61  & 70  & 1.038  & all & 9 & {\citeyear{2015A&A...583A.115B}}\\
V* SZ Hya &  & 138.4534  & -9.3191  & kA1hA7 & 102  & 72  & 2.647  & all & 3 & {\citeyear{2015A&A...583A.115B}}\\
V* U Lep &  & 74.0749  & -21.2172  & A8/F2(I) & 75  & 64  & 1.614  & all & 3 & {\citeyear{2015A&A...583A.115B}}\\
V* V1005 Ori &  & 74.8951  & 1.7835  & M0Ve & 55  & 180  & 0.093  & metal & 1 & \citeyear{1980ApJ...240..567V}\\
V* V1356 Ori &  & 92.1005  & 13.9466  & A0V & 2664  & 66  & 395.313  & all & 6 & {\citeyear{2006A&A...450..777B}},{\citeyear{2015A&A...583A.115B}}\\
\hline
		
	\end{tabular}
	
	\ec
\end{table}

\end{document}